\documentclass[12pt,letterpaper]{article}
\usepackage[a4paper, total={7in, 10in}]{geometry}
\date{}

\usepackage{amsmath,amssymb,mathtools,bbm}
\usepackage{amsthm}
\usepackage{mathrsfs}
\usepackage{stackrel}
\usepackage{nicefrac}
\usepackage[normalem]{ulem}

 \theoremstyle{remark}	

  \theoremstyle{remark}	
 
 \theoremstyle{remark}	

 \theoremstyle{remark}	

 \theoremstyle{remark} 

 \theoremstyle{remark} 
\newtheorem{remark}{Remark}
 \theoremstyle{remark}

\usepackage{comment}

\usepackage{helvet}

\usepackage{physics}

\usepackage{hyperref}
\hypersetup{colorlinks=true,linkcolor=blue,citecolor=blue,urlcolor=blue}

\usepackage{graphicx}
\usepackage{tikz}
\usetikzlibrary{arrows.meta, positioning}
\usepackage{float}

\usepackage[super,comma,sort&compress]  
   {natbib}
\usepackage[right]{lineno} 

\newcommand{\I}{\mathbbm{1}}

\newcommand{\hbin}{h_2}
\newcommand{\hk}{h_k}

\usepackage{authblk}
\usepackage{orcidlink} 
\author[1,2,6,\orcidlink{0009-0008-5109-1922}]{Amit Kam}
\author[3,6,\orcidlink{0000-0003-0594-6236},*]{Kfir Sulimany}
\author[3,4,\orcidlink{0000-0003-0167-3402}]{Shai Tsesses}
\author[2,5,7,\orcidlink{0000-0002-3259-6094},**]{Uzi Pereg}

\affil[1]{Department of Physics, Technion - Israel Institute of Technology, Haifa 32000, Israel}
\affil[2]{Helen Diller Quantum Center, Technion - Israel Institute of Technology, Haifa 32000, Israel}
\affil[3]{Research Laboratory of Electronics, Massachusetts Institute of Technology, Cambridge, MA 02139, USA}
\affil[4]{MIT-Harvard Center for Ultracold Atoms, Massachusetts Institute of Technology, Cambridge, MA 02139, USA}
\affil[5]{Andrew and Erna Viterbi Department of Electrical \& Computer Engineering, Technion - Israel Institute of Technology, Haifa 32000, Israel}
\affil[6]{These authors contributed equally}
\affil[7]{Lead contact}

\affil[*]{Correspondence: \href{mailto:kfir@mit.edu}{kfir@mit.edu} (K.S.)}
\affil[**]{Correspondence: \href{mailto:uzipereg@technion.ac.il}{uzipereg@technion.ac.il} (U.P.)}

\makeatletter
\renewcommand{\maketitle}{\bgroup\setlength{\parindent}{0pt}
\begin{flushleft}
  {\Large\LARGE\bfseries \@title \par}
  \vspace{0.5em}
  \@author
\end{flushleft}\egroup}
\makeatother

\title{Reduced State Embedding for Quantum Cryptography}

\begin{document}

\maketitle

\section*{SUMMARY}

Encoding in a high-dimensional Hilbert space improves noise resilience in quantum information processing. This approach, however, may result in cross-mode coupling and detection complexities, thereby reducing quantum cryptography performance. This fundamental trade-off between correctness and secrecy motivates the search for quantum high-dimensional encoding  approaches for cryptography. 
Here, we introduce state embeddings that use a $k$-symbol subset within a $d$-dimensional Hilbert space, tailored to the channel’s error structure.  
We demonstrate the advantage of our scheme in realistic quantum channels, producing a higher secure key rate. We validate our approach using a $d=25$ quantum key distribution (QKD) experimental data, derive closed-form expressions for the key rate and threshold, and determine the optimum at $k=5$. These findings advance high-dimensional QKD and pave the way to error-correction and modulation for quantum cryptography.

\section*{KEYWORDS}

Quantum Cryptography, Reduced State Embedding, Quantum Key Distribution,  BB84, High-dimensional Encoding, Quantum Modulation.

\section{INTRODUCTION}

Quantum key distribution (QKD) provides a provably secure means of sharing encryption keys between two remote parties by exploiting the fundamental principles of quantum mechanics, rather than computational assumptions~\cite{ekert_quantum_1991,scarani_security_2009,pereg2021key,lederman2024secure, berta2010uncertainty}. Since the introduction of the BB84 protocol~\cite{bennett_quantum_1984}, the study and implementation of QKD have developed into a vibrant research field~\cite{cao_evolution_2022}. QKD protocols have been demonstrated over long-distance optical fibers~\cite{wang_twin-field_2022}, satellite-to-ground links~\cite{liao_satellite_2017}, and free-space channels~\cite{xu_secure_2020,schmitt-manderbach_experimental_2007, bloom2025decoy}. These advances illustrate its potential for real-world deployment. Nevertheless,  achievable key rates remain strongly limited by loss, channel noise, and detector imperfections~\cite{li_efficient_2008, pirandola2017fundamental, diamanti2016practical}. Overcoming these bottlenecks is essential for extending QKD to global scales.

A promising route is to employ high-dimensional Hilbert spaces, where a state of light encodes not just a single qubit state, but a higher-dimensional 
state~\cite{cerf_security_2002,erhard_twisted_2018,erhard_advances_2020}. High-dimensional QKD increases the information capacity per detected photon and raises the tolerable error rate threshold~\cite{sheridan_security_2010}. Experimental demonstrations have employed various degrees of freedom, including spatial modes~\cite{scarfe_spatial_2025,walborn_quantum_2006,etcheverry_quantum_2013,mirhosseini_high-dimensional_2015,
sit_high-dimensional_2017,bouchard_high-dimensional_2017,bouchard_experimental_2018,cozzolino_orbital_2019,tentrup_large-alphabet_2019,zhou_using_2019,otte_high-dimensional_2020,
hu_pathways_2021,ortega_experimental_2021,stasiuk_high-dimensional_2023,halevi_high-dimensional_2024,nemirovsky-levy_nonlinear_2025, meyer2025analogy}, time-bin encoding~\cite{islam_provably_2017,lee_large-alphabet_2019,vagniluca_efficient_2020,ikuta_scalable_2022,chapman_hyperentangled_2022,sulimany_high-dimensional_2025, scarfe2025high}
 and time-energy entanglement~\cite{alikhan_large-alphabet_2007,mower_high-dimensional_2013,lee_entanglement-based_2014,zhong_photon-efficient_2015,liu_energy-time_2019,bouchard_achieving_2021,liu_high-dimensional_2023,bulla_nonlocal_2023,chang_large-alphabet_2023, tagliavacche2025frequency}. 

While high-dimensional encoding offers clear theoretical advantages, the direct approach of employing $d$ states of a $d$-dimensional Hilbert space in two mutually unbiased bases (MUBs) is experimentally challenging. As $d$ increases, state preparation and mode control become increasingly complex and costly~\cite{zhang_large-scale_2018};
interferometric stability becomes harder to maintain; and detection requires high efficiency across many parallel channels. 
Consequently, high-dimensional encoding can also increase the eavesdropper’s potential knowledge, hence the net security gain does not necessarily grow with the dimension~\cite{scarani_security_2009,sheridan_security_2010,shor2000simple}. In practice, imperfections such as mode mismatch, cross-talk, and detector noise accumulate with dimension, often negating the theoretical advantages and even reducing the secure key rate~\cite{renner_security_2008,bouchard_achieving_2021,lederman2024semantic}. 

These limitations suggest that the practical advantage of large Hilbert spaces is realized by encoding within a carefully chosen subspace and, crucially, adapting both the subspace size $k$ and the specific states to the channel's error structure. Standard QKD protocols delegate the error correction to the post-processing stage, while the quantum communication stage involves neither modulation nor error correction. Recent studies~\cite{doda2021quantum,kanitschar2025practical} have proposed comparable high‑dimensional encodings and analyzed a reduction during the public discussion in the classical post-processing stage, after the quantum transmission has concluded.

Here, we introduce  a strategy of state reduction within a high-dimensional embedding and provide experimental validation.
Rather than employing the full set of $d$ orthogonal states, our modulation scheme encodes information in a smaller subset of $k$ states within the same Hilbert space, where  $k<d$. This reduction is incorporated within the modulation of the quantum signal itself. We adopt a physical‑layer encoding approach, optimizing the encoding for realistic channel models. Although each transmission carries less information, we show that such a reduced-state protocol can outperform the full $d$-state protocol in robustness to noise and even achieve a higher secure key rate. We analyze three channel models that represent common implementations of QKD: depolarizing, modulo, and block-biased channels. Optimizing the signal sets reveals that fundamentally different encoding strategies are optimal for different noise models.

Our theoretical analysis of  the Devetak-Winter rate~\cite{devetak_distillation_2005} for each noise model yields closed-form expressions for the secure key rate, error threshold, and sifting efficiency,  and establishes noise-dependent interior optima for $k<d$. 
We validate our approach using a $d=25$ dimensional QKD system based on spatially entangled photon pairs~\cite{lib_processing_2022,lib_high-dimensional_2025}. By varying $k$, we study the dependence of the secure key rate on the reduced-state dimension for the block-biased channel and identify the optimal secure key rate at $k=5$ in agreement with the theoretical prediction. This encoding strategy  introduces aspects of state representation and physical-layer implementation into the primary stage of quantum transmission.

Our reduced-state embedding framework has potential applications beyond QKD, including blind quantum computation~\cite{broadbent2009universal}, quantum-secure multiparty deep learning~\cite{sulimany2024quantum}, and quantum direct secure communication~\cite{zhang2017quantum}. Moreover, our approach paves the way for quantum 
encoding schemes that balance capacity, security, robustness, and practicality.
\newpage

\section{RESULTS}

\subsection{Definitions}
\label{definitions}
We first introduce the key concepts for our state embedding scheme. 

\subsubsection{Basic Notation}
\label{Subsection:Basic_Notation}
 We use standard notation conventions\cite{wilde2013quantum}, as summarized below. 
A quantum state is described by a density operator $\rho$ on the Hilbert space $\mathcal{H}$. The dimensions are assumed to be finite. 
The probability distribution of a measurement outcome can be described in terms of a positive operator-valued measure (POVM), i.e. a set  of positive semidefinite operators $\{ \Gamma_j \}$, such that 
$\sum_j \Gamma_j=\I$, where $\I$ is the identity operator. 
According to the Born rule, if the system is in state $\rho$, then the probability of the measurement outcome $j$ is given by $\Pr(j)=\trace(\Gamma_j \rho)$.

The Shannon entropy with respect to a classical discrete random variable $X\sim p_X$ on a finite alphabet $\mathcal{X}$ is defined by 
$
H(p_X)=-\sum_{x\in\mathcal{X}} p_X(x) \log p_X(x)
$
under the convention of $0\log 0\equiv 0$.
The binary entropy for 
$p_X=(1-Q,Q)$ is denoted by 
$h_2(Q)=-(1-Q\log_2(1-Q)-Q\log_2 Q$. 
Similarly, for the $k$-ary distribution 
$p_X=\left(1-Q,\frac{Q}{k-1},\ldots,\frac{Q}{k-1}\right)$, we denote 
\begin{equation}
\label{eq:hk}
\begin{split}
\hk(Q)
&= -(1-Q)\log_2(1-Q) - Q\log_2\!\frac{Q}{k-1} \\
&= \hbin(Q) + Q\log_2(k-1).
\end{split}
\end{equation}
Define the quantum von Neumann entropy of the density operator $\rho$ as
$H(\rho) := -\trace[ \rho\log(\rho) ]$,
which is the same as the Shannon entropy associated with the eigenspectrum of $\rho$.
Consider the state of a bipartite system, composed of $A$ and $B$, on the tensor product $\mathcal{H}_A\otimes \mathcal{H}_B$ of the corresponding Hilbert spaces.
Given a bipartite state $\rho_{AB}$, define the conditional quantum entropy  by
$H(A|B)_{\rho}:=H(\rho_{AB})-H(\rho_B)$.

A classical-quantum state corresponding to a state ensemble $\{p_X(x), \rho_x\}_{x\in\mathcal{X}}$ has the form
$\rho_{XB}=\sum_{x\in\mathcal{X}} p_X(x) \ketbra{x}_X\otimes \rho_x$
where $\{\ket x\}$ is an orthonormal basis for $X$ (a classical register that stores the value $x$). In this case,
$
H(B|X)_\rho = \sum_{x \in \mathcal{X}} p_X(x)\, H(\rho_x)
$.
This expression admits a clear interpretation as an average entropy conditioned on the classical variable \(X\). In contrast, the reverse quantity
$
H(X|B)_\rho = H(XB)_\rho - H(B)_\rho
$
does not admit an immediate interpretation as conditioning in the classical sense. Rather, it quantifies the uncertainty about \(X\) in the presence of quantum side information \(B\), and is operationally meaningful in tasks such as privacy amplification~\cite{renner_security_2008} and data compression~\cite{devetak2003classical} with quantum side information.

\subsubsection{Signal Sets}
\label{Subsection:Signals}
Denote the computational and conjugate bases of the overall input space by 
\begin{align}
\mathcal B_Z  &=\{\ket{0},\ldots,\ket{d-1}\}
\intertext{and}
\mathcal B_X  &=\{\ket{\mu_0},\ldots,\ket{\mu_{d-1}}\}
\end{align}
respectively. The bases $\mathcal B_Z$ and $\mathcal B_X$ are  mutually unbiased, i.e.\ $\abs{\braket{j}{\mu_\ell}}^2=\tfrac1d$ for all $j,\ell\in \{0,1,\ldots,d-1\}$.

Our scheme uses a limited set of signals in a subspace of dimension $k$, where $k<d$. We begin with the simple scheme of truncation.
The $Z$-basis signals are the first $k$ orthonormal states:
\begin{equation}
\mathsf S_Z=\{\ket{0},\ldots,\ket{k-1}\}.
\end{equation}
Similarly, the $X$-basis signals are
\begin{equation}
\mathsf S_X=\{\ket{\mu_0},\ldots,\ket{\mu_{k-1}}\}.
\end{equation}

More generally, one may use any orthonormal subsets of size $k$, 
\begin{align}
\mathscr S_Z&=\{\ket{\phi_0},\ldots,\ket{\phi_{k-1}}\}
\intertext{and}
\mathscr S_X&=\{\ket{\sigma_0},\ldots,\ket{\sigma_{k-1}}\}
\end{align}
such that $\abs{\braket{\phi_j}{\sigma_\ell}}^2=\tfrac1d$ for all $j,\ell\in \{0,1,\ldots,k-1\}$.

For example, for $d=6$ and $k=3$, the simple $Z$-basis subset is
$\mathsf{S}_Z=\{
\ket0,
\ket1,
\ket2
\}$. Whereas,
one may choose instead the following $Z$-basis subset:
$\mathscr{S}_Z=\{
\ket{\phi_0}=\ket0,
\ket{\phi_1}=\ket2,
\ket{\phi_2}=\ket4\}$.
Later, in Section~\ref{Subsection:Mod}, we will see a noise model where the latter choice provides higher performance. 

The subsets notation is summarized in  Table~\ref{tab:notation_1}.

The  mutually unbiased pair $(\mathscr S_Z, \mathscr S_X)$ is hereby referred to as the state encoding. 

\subsubsection{Reduced state embedded QKD Protocol}
\label{Subsection:Protocol}
We incorporate the signal sets above within the BB84-QKD protocol as described below~\cite{bennett_quantum_1984,pirandola_symmetric_2008,wolf2021quantum}. 

\medskip
\noindent\textbf{Encoding.}
In each round, Alice picks a basis $b\in\{Z,X\}$ and a symbol $x\in\{0,\ldots,k-1\}$ uniformly at random,  and sends $\ket{\psi_{b,x}}=
 \ket x$ if $b=Z$, and $\ket{\psi_{b,x}}=
 \ket {\mu_x}$ if $b=X$.

\medskip
\noindent\textbf{Measurement and sifting.}
In each round, Bob independently and uniformly chooses a random measurement basis $b' \in \{Z,X\}$, without knowing Alice’s choice $b$, and applies the $(k{+}1)$-outcome positive operator-valued measure (POVM)
\begin{equation}
\label{eq:POVM}
\begin{split}
\Pi_{b',x} &= \ketbra{\psi_{b',x}}, \quad x=0,\dots,k-1,\\
\Pi_{b',\perp} &= \I - \sum_{x=0}^{k-1} \Pi_{b',x}.
\end{split}
\end{equation}
See Table~\ref{tab:notation_1}.
Outcomes $x\in\{0,\ldots,k-1\}$ are conclusive (``kept''); $\perp$ is inconclusive (``discarded'').
We denote the probability that Bob obtains a conclusive measurement (``kept") outcome for a matching round by 
$
\alpha
$. 
Bob informs Alice that he has received and measured all photons. Then, Alice and Bob publicly compare their basis choices and keep only the rounds for which $b'=b$ and the measurement has been conclusive. Upon sifting, the raw key consists of $k$-ary symbols, resulting from conclusive and basis-matched events.

\medskip
\noindent\textbf{Error Estimation.}
Alice and Bob disclose a random test set, i.e., a substring of the sifted key.
They compute the quantum dit error rate $Q$, i.e., the fraction of rounds in which Alice and Bob's outcomes do not agree.
The error rate provides a bound on Eve’s potential information.

\medskip
\noindent\textbf{Security Check and Post-Processing.}
If the error rate is below the security threshold, Alice and Bob proceed with the post-processing steps of Information Reconciliation and Privacy Amplification. 
The output is ideally a final, secret and error-free key (binary or $k$-ary, depending on the state encoding).

\medskip

\subsection{Performance Analysis}
\label{sec:performance}
To analyze the performance of reduced-state embedding in realistic scenarios, we examine its behavior under three representative noise models. First, we consider the \textbf{depolarization channel} $\mathfrak D(\rho)$, the canonical symmetric noise model that uniformly randomizes the state across the full Hilbert space. Next, we study the \textbf{modulo channel} $\mathfrak M(\rho)$, a structured noise where errors correspond to cyclic nearest-neighbor flips. Finally, we analyze the \textbf{block-bias channel} $\mathfrak B(\rho)$, which captures noise with preferential correlations inside contiguous subsets of states. Through these channel models, we show how reduced‑state embedding exploits the enlarged Hilbert space.

Following the sifting procedure, Alice and Bob only keep matching rounds, in which they use the same basis $b\in\{X,Z\}$.
Bob's measurement is conclusive only if the outcome falls within the signal subspace, $\mathrm{span}(\mathscr{S}_b)$.
Recall that we denote the probability that Bob obtains a conclusive measurement (``kept") outcome for a matching round by 
$
\alpha
$, and   the quantum dit error rate by $Q$, which results from the parameter estimation stage.
In the upcoming analysis, we characterize the relations between these quantities and the channel parameters.
In particular, we use $
\alpha_{\mathfrak D}
$, $
\alpha_{\mathfrak M}
$, and
$
\alpha_{\mathfrak B}
$
for the kept probability that is associated with the depolarizing, modulo, and block-biased channels, respectively.
Similarly, we use the respective notations of 
$
Q_{\mathfrak D}
$, $
Q_{\mathfrak M}
$, and 
$
Q_{\mathfrak B}
$, for the error rate. The channel-specific notations is summarized in  Table~\ref{tab:notation_2}.

Furthermore, denote the rate of final key bits per sifted symbol by
$R_{\mathrm{per\text{-}sifted\text{-}symbol}}$, i.e., the number of final key bits divided by the length of the sifted string.
Denote the rate of final key bits per signal by
$R_{\text{per-signal}}$, i.e., the number of final key bits divided by the overall number of transmissions.

\subsubsection{Depolarizing Channel}%
\label{Subsection:Dep}

We begin with the depolarizing channel, which provides a canonical model of symmetric noise in high-dimensional QKD, and thus serves as the natural starting point for our reduced-state embedding derivation. 

\begin{figure*}[!ht]
    \centering
    \includegraphics[width=\textwidth]{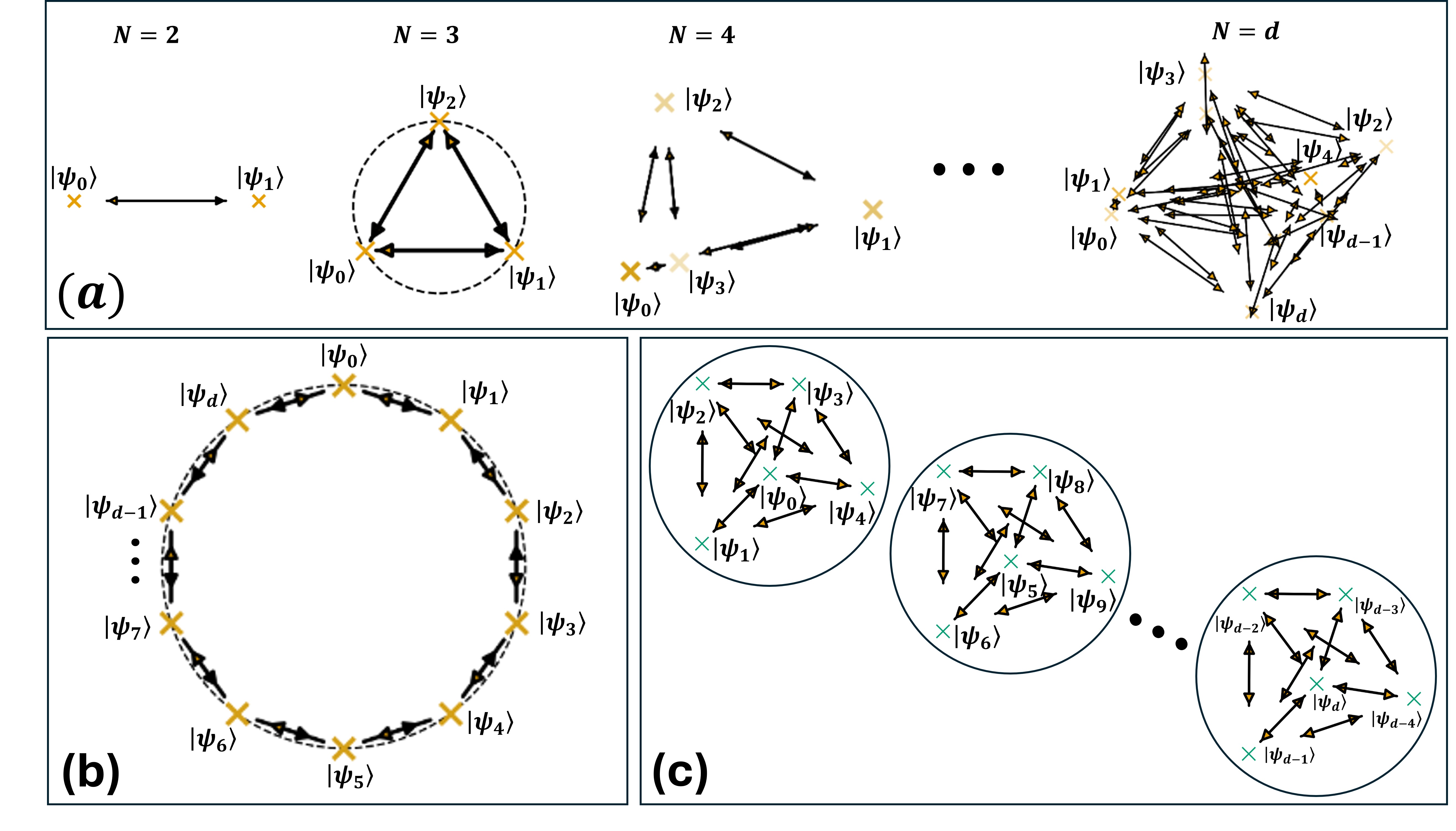}
    \caption{\textbf{Conceptual visualizations of the noisy channels.} The states are represented by nodes in a graph denote by ‘X', where the arrows indicate the connections or transitions between states induced by the corresponding noise model. 
    \textbf{(a) Depolarizing channel (Section~\ref{Subsection:Dep}).} 
    In the depolarization model, each state is equally distant from every other state, as every pair has the same transition probability.  The states sit at the vertices of a regular simplex (e.g., triangle for \(d=3\), tetrahedron for \(d=4\)), hence every pairwise distance is identical.
    \textbf{(b) Modulo channel (Section~\ref{Subsection:Mod}).} States are arranged at equal spacing on a ring. Transition is  only possible between two nearest-neighbors \(i\pm 1 \pmod d\).  
    \textbf{(c) Block-bias channel (Section~\ref{Subsection:block}).} 
    Illustrated here is an example of a state space partitioned into disjoint five-state blocks. Within each block, all-to-all transitions occur with equal probability, while only weak coupling is present between different blocks.}
    \label{fig:statesvis}
\end{figure*}

\paragraph{Model description} Consider a depolarizing channel model. The channel transformation replaces an input state by the maximally mixed state, with probability $\varepsilon$:
\begin{equation}
\label{eq:depolarizing}
\mathfrak D (\rho)=(1-\varepsilon)\rho+\frac{\varepsilon}{d}\,\I_d,\qquad 0\le \varepsilon\le 1.
\end{equation}
Figure~\ref{fig:statesvis}(a) illustrates the noise model.
In particular,
 for every pure input signal $\ket{\psi}$, the output state is given by 
\begin{equation}
\label{eq:rhoout}
\rho_{\text{out}}=(1-\varepsilon)\ket{\psi}\!\bra{\psi}+\frac{\varepsilon}{d}\,\I_d.
\end{equation}

\paragraph{Kept probability and dit error for a general encoding.}

We now derive an expression for the ``kept" probability  
$
\alpha_{\mathfrak D} 
$ in terms of the depolarization probability $\varepsilon$, the signal subset size $k$, and the Hilbert space dimension $d$. 
Let $P_b=\sum_{x=0}^{k-1}\ketbra{\psi_{b,x}}$ be the projector onto the signal subspace, for a given basis $b$ (see Section~\ref{Subsection:Protocol}, Table~\ref{tab:notation_1} and Table ~\ref{tab:notation_2}). Notice that the projector is of rank $\trace(P_b)=k$.
For a given transmitted signal $\ket{\psi_{b,x}}$, the conditional kept probability is then:
\begin{align}
\Pr[\text{kept}\mid \text{basis match}, b,x]
& = \tr\!\big(\rho_{\text{out}} P_b\big)\nonumber\\
&= (1-\varepsilon)\,\tr(\ket{\psi_{b,x}}\!\bra{\psi_{b,x}} P_b)
   + \frac{\varepsilon}{d}\,\tr(P_b) \nonumber\\
&= (1-\varepsilon)\cdot 1 + \frac{\varepsilon}{d}\cdot k
= (1-\varepsilon)+\frac{k\,\varepsilon}{d}.
\end{align}
for $b\in\{X,Z\}$ and $x\in\{0,1,\ldots,k-1\}$.
Hence, the kept probability is:
\begin{align}
\alpha_{\mathfrak D} 
&
= (1-\varepsilon)+\frac{k\,\varepsilon}{d}.
\label{eq:kappa}
\end{align}
Within the kept subspace, the conditional confusion matrix is symmetric:
\begin{equation}
\Pr[y \mid x \wedge \text{kept}] =
\begin{cases}
\dfrac{(1-\varepsilon)+\tfrac{\varepsilon}{d}}{\alpha_{\mathfrak D} }, & \text{if } y = x, \\[8pt]
\dfrac{\tfrac{\varepsilon}{d}}{\alpha_{\mathfrak D} }, & \text{if } y \neq x.
\end{cases}
\label{eq:prob-cases}
\end{equation}
for all $x,y\in\{0,1,\ldots,k-1\}$.

Therefore, the $k$-ary error rate among kept-events (dit error) is
\begin{equation}
\label{eq:Qkd}
Q_{\mathfrak D} 
= \frac{(k-1)\,\frac{\varepsilon}{d}}{(1-\varepsilon)+\frac{k\,\varepsilon}{d}}
= \frac{(k-1)\,\varepsilon}{d(1-\varepsilon)+k\,\varepsilon}\,.
\end{equation}

\paragraph{Optimal encoding.} We use a state encoding that exploits only $k$ out of $d$ degrees of freedom.
Due to the symmetry of the depolarization noise model, it suffices to consider the simple truncation encoding scheme $(\mathsf S_Z, \mathsf S_X)$ from Section~\ref{Subsection:Signals}.

\paragraph{Key rate and error thresholds.}
Next, we derive an expression for the key rate in terms of the signal subset size $k$ and the error rate $Q_{\mathfrak D}$ (see Section~\ref{Subsection:Protocol}, Table~\ref{tab:notation_1} and Table ~\ref{tab:notation_2}). 
For the analysis, suppose that upon the sifting stage, Eve's interference has resulted in a classical-quantum state
$\rho_{V_A V_B E}$, where $V_A$ and $V_B$ stand for Alice and Bob's sifted keys, respectively, and $E$ is Eve's quantum side information. 
For a two-basis protocol with symmetric errors and key extracted from such raw keys, the asymptotic Devetak-Winter bound~\cite{devetak_distillation_2005} per sifted symbol is:
\begin{equation}
\label{eq:DW}
\begin{split}
R_{\mathrm{per\text{-}sifted\text{-}symbol}}
\ \ge\ &\, H(V_A|E)-H(V_A|V_B) \\
=\ &\, \log_2 k - h_k(Q_{Z,\mathfrak D}) - h_k(Q_{X,\mathfrak D}) \\
=\ &\, \log_2 k - 2\,h_k\!\big(Q_{\mathfrak D} )\,,
\end{split}
\end{equation}
where $\hk(\cdot)$ is the Shannon entropy of the $k$-ary symmetric channel with error $Q$ (see \eqref{eq:hk} in Section~\ref{Subsection:Basic_Notation}).

A positive key rate sets the threshold at 
\begin{equation}
\label{eq:Qstar-eq}
\hk\!\big(Q_{\mathfrak D}^{\text{th}}  \big)=\tfrac12\log_2 k,
\end{equation}
which defines the observed (kept, matched) dit-error threshold $Q_{\mathfrak D}^{\text{th}}$, independent of $d$.

Numerical calculation yields the following threshold values, a standard result which we next generalize:
\[
\begin{array}{c|cccccccc}
k & 2 & 3 & 4 & 5& 6 \\\hline
Q_{\mathfrak D}^{\text{th}}   & 0.1100 & 0.1595 & 0.1893 & 0.2099 & 0.2252 
\end{array}
\]
(rounded to $4$ significant digits).

\begin{figure*}[!ht]
    \centering
    \includegraphics[width=1\columnwidth]{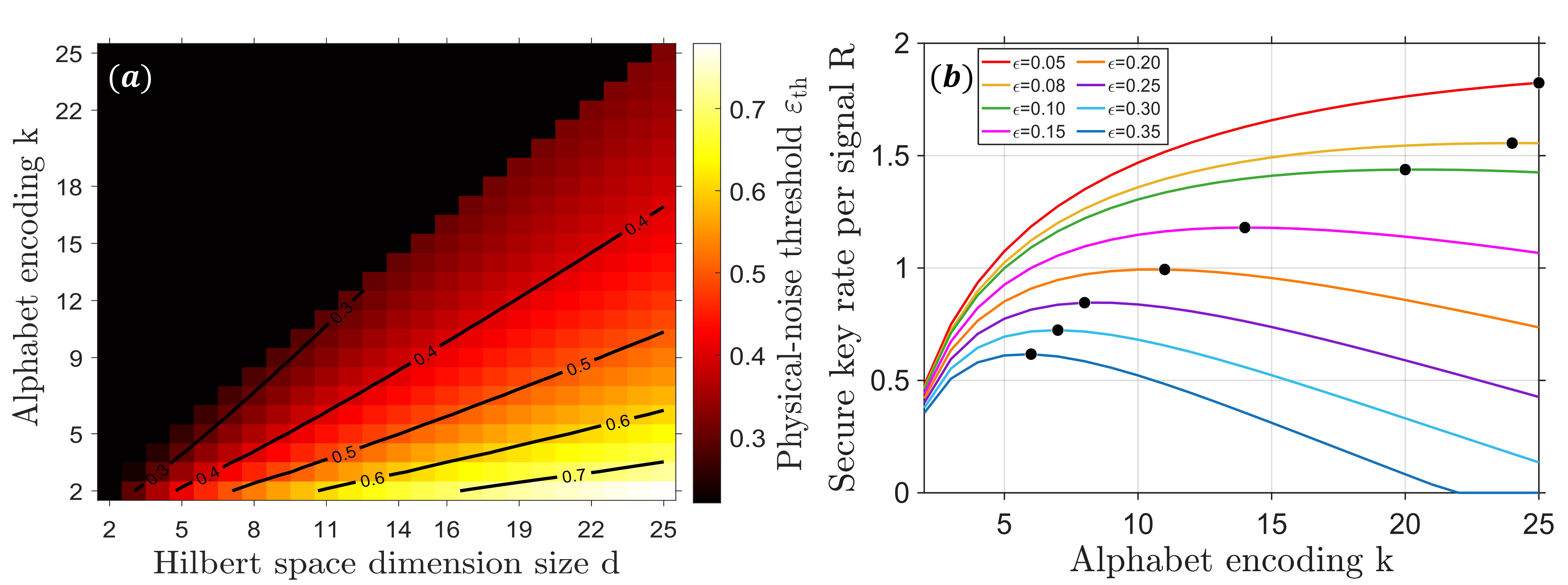}
        \caption{\textbf{Physical-noise threshold and secure key rate for the depolarizing channel.} 
\textbf{(a) Physical-noise threshold $\varepsilon_{\mathfrak{D}}^{\mathrm{th}} $ for different values of $d$}. The heatmap shows the threshold of the tolerable depolarization probability, for a positive Devetak--Winter key rate, as a function of the signal-set size $k$ and the space dimension $d$. 
\textbf{(b) Secure key rate $R$ as a function of signal-set size $k$, for a fixed  dimension $d=25$.} Each curve corresponds to a different physical noise parameter $\varepsilon$. For every $\varepsilon$, a black dot marks the optimal signal-set size, which  maximizes the secret key rate. The plot highlights the trade-off between increasing signal-set size and noise accumulation. Initially, as $k$ increases,  the key rate increases as well. For larger $k$, however, noise accumulation may suppress 
performance. Notably, for $\varepsilon \approx 0.083$, the optimal performance occurs when the signal-set size is strictly smaller than the space dimension, i.e., $k<25$. This confirms that encoding into a reduced subspace is preferable to using the full Hilbert space dimension.}
    \label{fig:pth_vs_k_dip}
\end{figure*}

Combining \eqref{eq:Qkd} with $Q=Q_{\mathfrak D}^{\text{th}}  $ and solving for $\varepsilon$, we obtain the following relation for the threshold of the depolarization probability:
\begin{equation}
\label{eq:pth}
\varepsilon^{\text{th}}_{\mathfrak D} 
=\frac{d\,Q_{\mathfrak D}^{th}  }{(k-1)+Q_{\mathfrak D}^{th}  \,(d-k)}
\end{equation}
which monotonically increases in $d$, for a fixed $k$. See Figure~\ref{fig:pth_vs_k_dip}(a). Note that $\varepsilon^{\text{th}}_{\mathfrak D} $ tends to $1$ as $d\to\infty$.

 \begin{remark}
For $k=2$, we recover the familiar qubit error results in a $d$-dimensional space: the error rate threshold is $\approx 11.0\%$ for all $d$, while the physical depolarizing threshold $\varepsilon^{\text{th}}_{\mathfrak D}(2,d)$ increases monotonically with $d$ and approaches $1$ as $d\to\infty$. 
 \end{remark}

The corresponding threshold for the kept-event fraction  follows from \eqref{eq:kappa}:
\begin{equation}
\label{eq:kappa-th}
\begin{split}
\alpha^{\text{th}}_{\mathfrak D} 
&= 1 - \varepsilon^{\text{th}}_{\mathfrak D}\!\left(1 - \tfrac{k}{d}\right) \\[6pt]
&= \frac{k-1}{(k-1)+Q_{\mathfrak D}^{th}  \,(d-k)} 
\end{split}
\end{equation}
which tends to zero as $d\to\infty$.

We observe that our embedding of a $k$-limited signal set (for $k<d$) increases robustness to physical depolarizing noise, but reduces the sifted-throughput (kept probability).

\begin{remark}
As the number of signals, $k$, becomes larger, while the space dimension $d$ remains fixed, the observed error threshold $Q_{\mathfrak D}^{th}  $ increases. On the other hand, the physical threshold, i.e., the tolerable depolarization probability $\varepsilon^{\text{th}}_{\mathfrak D} $, decreases, since a larger fraction of the depolarization noise lies inside the $k$-dimensional kept subspace.
\end{remark}

For a uniform basis selection, the sifted fraction per signal is $\tfrac12\,\alpha_{\mathfrak D} $. The ideal asymptotic Devetak-Winter secret bits per signal therefore obeys
\begin{equation}
R_{\text{per-signal}} \ge \frac12\alpha_{\mathfrak D} \Big[\log_2 k - 2\,\hk\!\big(Q_{\mathfrak D} \big)\Big].
\end{equation}
(see \eqref{eq:kappa} and \eqref{eq:Qkd} for the definition of $\alpha_{\mathfrak D}  $ and $Q_{\mathfrak D}  $, respectively).

\paragraph{Advantage of reduced state embedding}
Figure~\ref{fig:pth_vs_k_dip}(b) highlights the advantage of our reduced state embedding scheme for the depolarizing channel. The figure depicts the key rate $R_{\text{per-signal}}$ as a function of the signal-set size $k$, for a fixed  dimension of  $d=25$. 
The plot highlights the trade-off between increasing signal-set size and noise accumulation. Initially, as $k$ increases,  the key rate increases as well. For larger $k$, however, noise accumulation may suppress 
performance. Notably, for $\varepsilon \approx 0.083$, the optimal performance occurs when the signal-set size is strictly smaller than the space dimension, i.e., $k<25$. Remarkably, this confirms that encoding into a reduced subspace is preferable to using the full Hilbert space dimension.

\subsubsection{Modulo Channel}
\label{Subsection:Mod} 
\begin{figure*}[!tb]
    \centering
    \includegraphics[width=1\columnwidth]{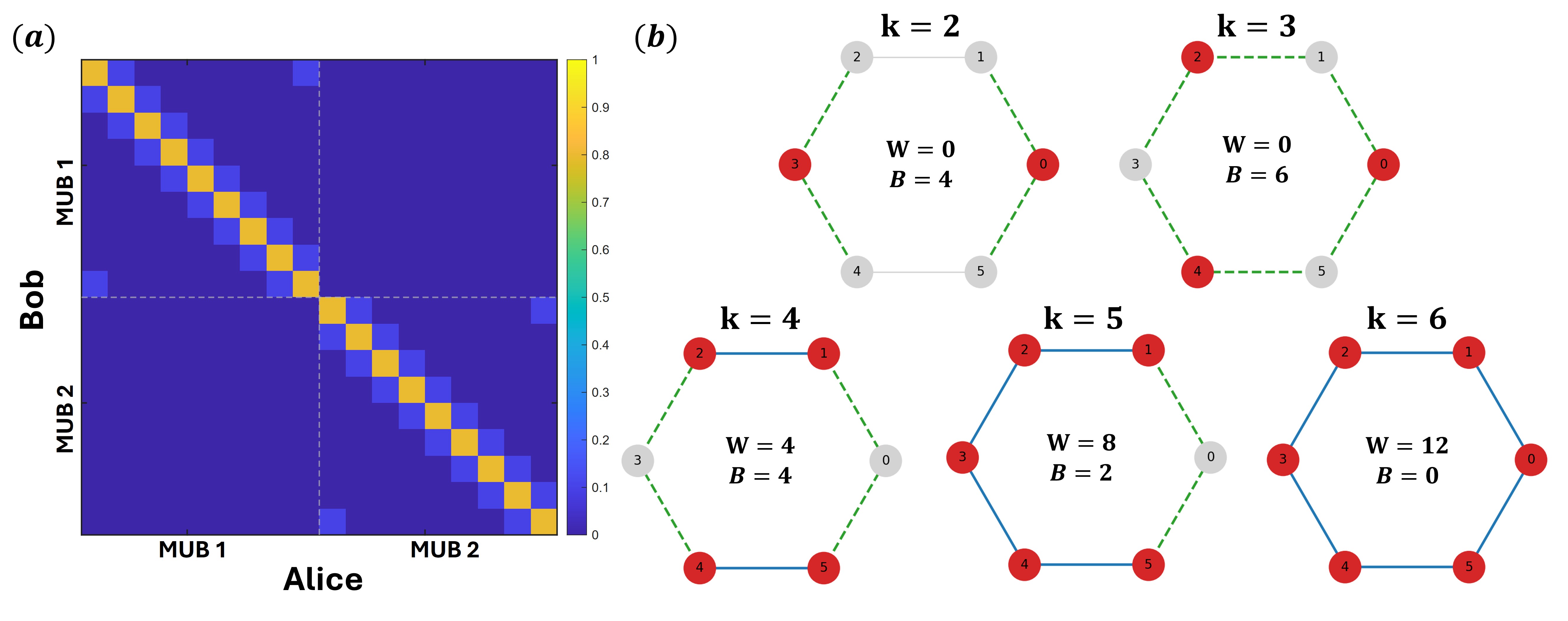}
    \caption{\textbf{Encoding a signal set of size $k$ on the cycle graph $C_6$.} \textbf{(a) Confusion matrix of the modulo channel for two mutually unbiased bases.} The main diagonal corresponds to correct outcomes, while the dominant errors are restricted to nearest neighbors within each basis, producing the structure of a cycle-graph with periodic wrap-around. Cross-basis terms are negligible.
    \textbf{(b) Optimal encoding for $k=2,\dots,6$.} 
   Red nodes represent chosen states in the subset $\mathscr S_b$, corresponding to basis $b$. Blue edges indicate internal adjacencies in $W(\cdot)$
   (``confusions") that generate errors within the kept set, 
and green dashed edges are boundary adjacencies in $B(\cdot)$ that lead to inconclusive outcomes. 
For $k\leq 3$, the encoding removes all internal adjacencies, i.e., $W=0$. Whereas, for $k>3$, some adjacencies are unavoidable, 
causing a trade-off between kept probability $\alpha$ and dit error rate $Q$.}
    \label{fig:coding_diff}
\end{figure*}

Next, we implement the reduced-state embedding within the modulo noise channel, a practical model for multicore optical fiber QKD systems \cite{ding_high-dimensional_2017,dalio_path-encoded_2021,zhou_high-fidelity_2021,zahidy_practical_2024}.
\paragraph{Model description} Consider the random-unitary channel,
\begin{equation}
\label{eq:NNchannel}
\mathfrak M (\rho) \;=\; (1-2\varepsilon)\,\rho \;+\; \varepsilon\,X\rho X^\dagger \;+\; \varepsilon\,X^\dagger \rho X,
\end{equation}
for $0\le \varepsilon\le \tfrac12$,
where $X$ is the qudit shift operator: $X\ket{j}=\ket{j{+}1 \bmod d}$, hence $X^\dagger\ket{j}=\ket{j{-}1 \bmod d}$.
The model describes nearest-neighbor hopping on a cyclic mod-$d$ state space, see Figure~\ref{fig:coding_diff}(a) for the confusion matrix of the channel.

Let $C_d$ denote the cycle graph, the vertices of which are $\{0,\ldots,d{-}1\}$, with edges between $j$ and $j\pm1$ mod $d$.

\paragraph{Kept probability and dit error for a general encoding.}
Recall that $\mathscr{S}_b$ denotes the signal set in our protocol, for each basis $b\in\{X,Z\}$ (see Section~\ref{Subsection:Signals}
and 
Table~\ref{tab:notation_1}).
On the cycle graph $C_d$, 
the signal subset $\mathscr S_b$ induces two directed counts:
\begin{equation}
\label{eq:W_B}
\begin{aligned}
W(\mathscr S_b) &:= \#\bigl\{\, (x \!\in\! \mathscr S_b) \to (x \pm 1 \!\in\! \mathscr S_b)\,\bigr\}, \\[6pt]
B(\mathscr S_b) &:= \#\bigl\{\, (x \!\in\! \mathscr S_b) \to (x \pm 1 \!\notin\! \mathscr S_b)\,\bigr\}.
\end{aligned}
\end{equation}
for $b\in\{X,Z\}$.
Intuitively, $W(\cdot)$ counts the internal nearest-neighbor adjacencies among chosen symbols (inducing kept errors), while $B(\cdot)$ counts the boundary adjacencies from chosen symbols into the discarded subspace (inducing inconclusiveness).
Each vertex has two neighbors, hence the identity
\begin{equation}
\label{eq:WBidentity}
W(\mathscr S_b)+B(\mathscr S_b)=2k
\end{equation}
where $2k$ is twice the signal-set size of our encoding space. 

To better illustrate the encoding notation introduced in Eq.~\eqref{eq:W_B}, consider the \(k=4\) encoding example shown in Fig.~\ref{fig:coding_diff}(b) of cycle graph \(C_{6}\). We choose the subset $\{1,2,4,5\}\subset C_6,$ where the vertex set of the cycle graph is $C_6=\{0,1,2,3,4,5\}$.
The corresponding edges are $\{(0,1), (1,2), (2,3), (3,4), (4,5), (5,0)\}$. The set \(W(\mathscr{S}_b)\) contains the internal adjacencies, $W(\mathscr{S}_b)=\{(1,2),(4,5)\}$. These edges correspond to nearest-neighbor confusions between kept states. Since we count directed confusions, each undirected edge contributes two transitions, namely $1\to2, 2\to1, 4\to5, 5\to4$. Therefore $W=4$. The set \(B(\mathscr{S}_b)\) contains the boundary adjacencies, namely edges with one endpoint in \(\mathscr{S}_b\) and the other in its complement. Thus, $B(\mathscr{S}_b)=\{(0,1),(2,3),(3,4),(5,0)\}$.
Counting the directed transitions gives $1\to0,2\to3,4\to3,5\to0$. Hence, $B=4$.
In summary, for the subset \(\mathscr{S}_Z=\{\ket{1},\ket{2},\ket{4},\ket{5}\}\), the two internal edges \((1,2)\) and \((4,5)\) generate four directed within-set confusions, giving \(W=4\), while the four boundary edges \((0,1)\), \((2,3)\), \((3,4)\), and \((5,0)\) generate four directed exits from the kept set, giving \(B=4\).

We now derive an expression for the ``kept" probability  
$
\alpha_{\mathfrak M} 
$ in terms of the shift probability $\varepsilon$, internal and boundary nearest-neighbor adjacencies, $W(\cdot)$ and $B(\cdot)$,  the signal subset size $k$, and the Hilbert space dimension $d$ (see Section~\ref{Subsection:Protocol}, Table~\ref{tab:notation_1} and Table ~\ref{tab:notation_2}). 

If Alice sends $x\in \mathscr S_b$ and Bob measures in basis $b$, the probability to keep a given round is
\begin{align}
&\Pr[\text{kept}\mid \text{basis match}, b,x]
\nonumber\\
&
=1-\varepsilon\big(\mathbf 1_{x+1\notin \mathscr S_b}+\mathbf 1_{x-1\notin \mathscr S_b}\big).
\end{align}
Averaging uniformly over $x\in \{0,\ldots,k-1\}$ yields the kept-event probability:
\begin{equation}
\label{eq:kappa2}
\ \alpha_{b,\mathfrak M}=1-\frac{\varepsilon}{k}\,B(\mathscr S_b)
      \ =\ 1-2\varepsilon+\frac{\varepsilon}{k}\,W(\mathscr S_b).
\end{equation}
The $k$-ary symbol error (“dit error”) in basis $b$, conditioned on a kept and basis-matched round, is the probability that the detection outcome corresponds to a neighboring symbol within $\mathscr S_b$:
\begin{equation}
\label{eq:Qb2}
\ Q_{b,\mathfrak M}=
\frac{\tfrac{\varepsilon}{k}\,W(\mathscr S_b)}{\alpha_{b,\mathfrak M}(\mathscr S_b)}=
\frac{\tfrac{\varepsilon}{k}\,W(\mathscr S_b)}{\,1-2\varepsilon+\tfrac{\varepsilon}{k}\,W(\mathscr S_b)\,}.
\end{equation}
Equations \eqref{eq:kappa2}–\eqref{eq:Qb2} are analogous to the depolarizing case, except that there is an explicit dependence on the geometry of the chosen coding $\mathscr S_b$, through $W(\cdot)$ and $B(\cdot)$.

\paragraph{Optimal encoding.}
If $k\le \lfloor d/2\rfloor$, we may choose signal subsets $\mathscr S_Z,\mathscr S_X$ without internal adjacencies, i.e.\ $W(\mathscr S_Z)=W(\mathscr S_X)=0$. In the example in Figure~\ref{fig:coding_diff}(b) above, for $d=6$ and $k_{\text{opt}}=3$, we take $\mathscr{S}_Z=\{\ket0,\ket2,\ket4\}$
and $\mathscr{S}_X=\{\ket{\mu_0},\ket{\mu_2},\ket{\mu_4}\}$. As a result, there are no adjacencies, i.e., $W(\cdot)=0$.

\begin{figure*}[ht]
    \centering
    \includegraphics[width=1\columnwidth]{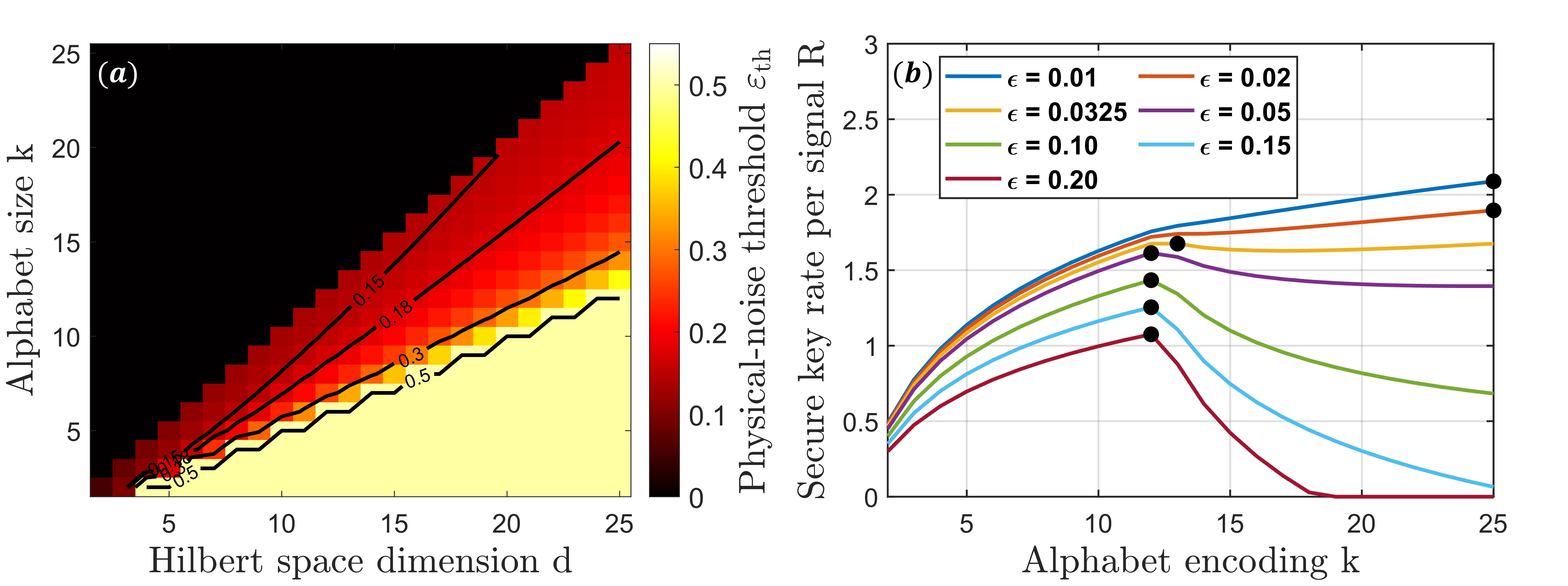}
        \caption{\textbf{Physical-noise threshold and secure key rate for modulo channel.} \textbf{(a) Heatmap of the physical-noise threshold $\varepsilon_{\mathfrak M}^{\mathrm{th}}$.} Results correspond to the evenly spaced encoding strategy. The triangular region $k \leq d$ marks valid encodings, with lighter colors indicating higher tolerance to physical noise. The plateau at $\varepsilon_{\mathfrak M}^{\mathrm{th}}=1/2$ identifies the maximal noise-tolerance regime, occurring whenever adjacent symbols can be completely avoided ($W=0$), so errors vanish and only inconclusive outcomes remain. \textbf{(b)  Secure key rate $R$ as a function of signal-set size $k$, for a fixed  dimension $d=25$.} Each curve corresponds to a different physical noise parameter $\varepsilon$. For every $\varepsilon$, a black dot marks the signal-set size that maximizes the  key rate. Notably, for $\varepsilon > 0.0325$, the optimal performance occurs at $k<d$, specifically $k_{\text{opt}}=\lfloor d/2\rfloor$, indicating that encoding into a reduced subspace is preferable to using the full Hilbert space dimension.}
    \label{fig:pth_vs_k_modulo}
\end{figure*}

Then,
\begin{equation}
\alpha_{b,\mathfrak M}=1-2\varepsilon,\qquad
Q_{b,\mathfrak M}=0.
\end{equation}
Thus, all neighbor flips are filtered out as inconclusive events.  
The threshold condition \eqref{eq:Qstar-eq} is saturated at $\varepsilon^{\text{th}}_{\mathfrak M} =\tfrac12$,
which represents the physical noise limit of the channel.  
The price is a vanishing kept rate $\alpha_{b,\mathfrak M}\to 0$ as $\varepsilon\to \tfrac12$. Thereby, while the secret key rate per kept symbol remains positive, the overall throughput tends to zero, see Figure~\ref{fig:pth_vs_k_modulo}(a).

\paragraph{Key rate and error thresholds.}
For the two-basis, $k$-ary protocol with symmetric sampling and one-way reconciliation, the Devetak-Winter lower bound reads
\begin{equation}
\label{eq:DW2}
R_{\mathrm{per\text{-}sifted\text{-}symbol}}\ \ge\ \log_2 k - h_k(Q_{Z,\mathfrak{M}}) - h_k(Q_{X,\mathfrak{M}}).
\end{equation}
A positive key rate requires
\begin{equation}
\label{eq:threshold_condition}
h_k\!\big(Q_{Z,\mathfrak{M}}\big)\ +\ h_k\!\big(Q_{X,\mathfrak{M}}\big)\ <\ \log_2 k
\end{equation}
where $h_k(\cdot)$ is as in 
\eqref{eq:hk}.

In a symmetric design, we use $W(\mathscr S_Z)=W(\mathscr S_X)=W$, hence
 $Q_{Z,\mathfrak M}=Q_{X,\mathfrak M}=:Q_{\mathfrak M}$. The condition~\eqref{eq:threshold_condition} becomes
\begin{align}
2\,h_k\!\big(Q_{\mathfrak M}\big) &< \log_2 k \notag \\[6pt]
\Rightarrow 
h_k\!\big(Q_{\mathfrak M}\big) &= \tfrac12 \log_2 k 
\quad \text{at threshold.}
\end{align}

Let $Q^{\text{th}}_{\mathfrak M}  $ be the unique solution of $h_k(Q^{\text{th}}_{\mathfrak M}  )=\tfrac12\log_2 k$.  
Using \eqref{eq:Qb2} and solving for $\varepsilon$ gives:
\begin{equation}
\label{eq:eps_threshold_symmetric}
\ \varepsilon_{\mathfrak M}^{\text{th}}=
\frac{Q_{\mathfrak M}^{\text{th}}  }{\,2Q_{\mathfrak M}^{\text{th}}  +\tfrac{W}{k}\big(1-Q_{\mathfrak M}^{\text{th}}  \big)}.
\end{equation}

For a symmetric design, the secret bits per signal rate therefore obeys
\begin{equation}
\label{eq:R_modulo_symmetric}
R_{\text{per-signal}}
\ \ge\
\tfrac12\,\alpha_{\mathfrak M}\,
\Big[\log_2 k - 2\,h_k\!\big(Q_{\mathfrak M}\big)\Big].
\end{equation}

\paragraph{Advantage of reduced state embedding}
Figure~\ref{fig:pth_vs_k_modulo}(b) highlights the advantage of our reduced state embedding scheme for the modulo channel. The figure depicts the key rate $R_{\text{per-signal}}$ as a function of the signal-set size $k$, for a fixed  dimension of  $d=25$. For $\varepsilon >0.0325$, the optimal  signal-set size is  $k_{\text{opt}}=\lfloor d/2\rfloor$. As in the depolarizing channel, this confirms once more that encoding into a reduced subspace is preferable.
Note that all  rates and parameters  (e.g., $\alpha_{\mathfrak M}  $, $Q_{\mathfrak M}$)  depend on the choice of signal subspaces 
$\mathscr{S}$.

\begin{remark}
In QKD systems based on time-bin qubit modulation, the physical channel is typically not cyclic. Specifically, there is no transition from $\ket{d-1}$ to $\ket{0}$, or vice versa.
In this case, the same derivation holds after replacing $W,B$ with the corresponding directed counts. The identities become $W+B=\sum_{x\in \mathscr S_b}\deg(x)$, hence \eqref{eq:kappa2}–\eqref{eq:Qb2} still hold. 
Edge effects matter only when the signal subset $\mathscr S_b$ includes the extremes $\ket{0}$, $\ket{d-1}$. For $k\ll d$ and well-separated coding, these extreme effects are negligible. 
\end{remark}

\subsubsection{Block-Biased Channel}
\label{Subsection:block}
The depolarizing and modulo channels capture two important extremes: completely symmetric noise and strictly nearest-neighbor errors. See Sections \ref{Subsection:Dep} and 
\ref{Subsection:Mod}, respectively.
Nonetheless, those models  do not necessarily reflect realistic behavior observed in  real high-dimensional QKD implementations. 

In practice, experimental imperfections often concentrate within subsets of modes, producing error patterns that are neither uniform across the Hilbert space, nor confined to cyclic adjacency. To account for such error patterns, we introduce the block-bias channel, a model in which noise preferentially acts within contiguous subsets (blocks) of states.
The block-biased channel arises naturally in multimode communication platforms, such as in multimode fibers \cite{ploschner2015seeing,sulimany2022all} and free-space  \cite{krenn2014generation,lib_high-dimensional_2025}, where coupling is strongest within mode subsets and the confusion matrix exhibits strong in-block errors and weak inter-block errors.

\paragraph{Model description} We explain the model of the block-biased channel below.
In accordance with our experimental validation in Section~\ref{sec:validation}, 
we consider a Hilbert space of dimension $d=s^2$, and partition the computational basis $\{\ket{j}\}_{j=0}^{d-1}$ into $s$ blocks of size $s$. A similar analysis  applies to an arbitrary block size $s$ that divides $d$. The modes interfere within each chosen measurement/encoding basis. Consequently, the effective noise is basis-conditioned which is block-biased in whichever basis $b$ is used.
We express the computational basis states as $\ket{j}=\ket{sm+r}$,
where $m$ is the block index, $m\in\{0,\dots,s-1\}$, and $r$ is the  index of states within each block, $r\in\{0,\dots,s-1\}$. 

If Bob measures in the basis $b$, then
 block depolarization  replaces the state of each block by the maximally mixed state on this block, weighted by its total population:
\begin{align}
\Phi_{\mathrm{block}}^{(b)}(\rho) 
  &= \sum_{m=0}^{s-1} \frac{U_b\Pi_m U_b^{\dagger} }{s}\,\Tr(U_b\Pi_mU_b^{\dagger} \rho),
  \label{Equation:Phi_b}
  \intertext{where $\Pi_m$ is the projector onto the subspace of block $m$:}
\Pi_m 
  &= \sum_{r=0}^{s-1}\ket{sm+r}\!\bra{sm+r}. 
\end{align}
and $U_b$ is a unitary that rotates $
\Pi_m$ to the measurement basis.
The map $\Phi_{\mathrm{block}}^{(b)}$ uniformly smears each signal across its block. 
Hence, the block-depolarizing map is
\begin{align}
\Lambda_{\mathrm{block}}^{(b)}=
(1-\varepsilon_1)\mathrm{id}+\varepsilon_1 \Phi_{\mathrm{block}}^{(b)}
\end{align}
where
$\mathrm{id}$ is the identity map on the $d$-dimensional space, and 
$\varepsilon_1$ is the probability of blockwise depolarization. 

Subsequently, 
a global depolarizing map acts as well:
\begin{align}
\Lambda_{\mathrm{global}}(\rho)=
(1-\varepsilon_2)\rho+\varepsilon_2 \frac{\I_d}{d}\,\Tr(\rho)
\end{align}
where $\varepsilon_2$ is the probability of global depolarization. 

\begin{figure*}[!ht]
    \centering
    \includegraphics[width=1\columnwidth]{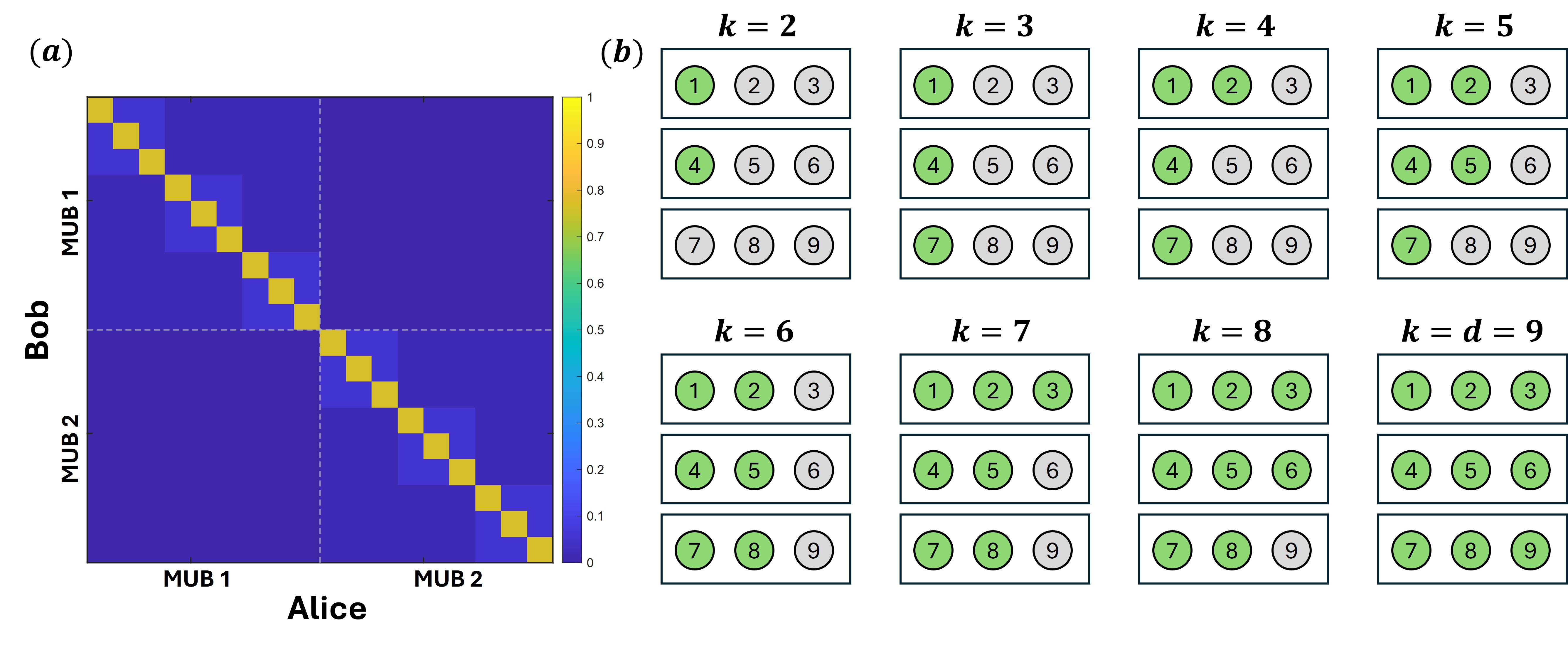}
    \caption{\textbf{Encoding a signal set of size \(k\) under a block-bias channel with block size \(s=3\).} \textbf{(a) Confusion matrix of the block-bias channel in two mutually unbiased bases (MUBs).} The dominant weight lies on the main diagonal, corresponding to correct transmission, while the leading errors are concentrated within \(3\times 3\) blocks, reflecting strong confusion among states belonging to the same block and weak mixing between different blocks. \textbf{(b) Optimal encoding}. Examples of the chosen subset \(\mathscr{S}_b\) for \(k=2,\dots,9\). The \(d=9\) states are partitioned into three blocks of size \(s=3\), shown as separate rows. Green circles denote the selected states in the subset \(\mathscr{S}_b\), while gray circles denote excluded states. As \(k\) increases, the encoding progressively fills the available blocks. This representation illustrates the trade-off imposed by the block-bias channel: distributing the chosen states across different blocks suppresses internal subset confusions, whereas selecting multiple states from the same block increases the probability of errors within the kept set.}
    \label{fig:coding_blockfig}
\end{figure*}

Overall, the block-biased channel is the composition:
\begin{equation}
\mathfrak B^{(b)} (\rho)=
(\Lambda_{\mathrm{global}}\circ
\Lambda_{\mathrm{block}}^{(b)})(\rho).
\end{equation}
See Figure~\ref{fig:coding_blockfig}(a) for the confusion matrix of the channel.
\paragraph{Kept probability and dit error for a general encoding.} We now express the ``kept"
probability  
$
\alpha_{\mathfrak B} 
$ in terms of the block- and global- depolarization probabilities, $\varepsilon_1$ and $\varepsilon_2$, block size $s$, signal subset size $k$, and  Hilbert space dimension $d$ (see Section~\ref{Subsection:Protocol}, Table~\ref{tab:notation_1} and Table ~\ref{tab:notation_2}). 

Fix the basis $b$ and the associated eigenbasis. Given an input state $\ket{m,r}$, the output of the block-biased channel is distributed over three types of outcomes: the state remains unchanged, the $s-1$ neighbors inside the same block are  depolarized, or the $d-s$ states in other blocks are  depolarized.
The respective probabilities of these three events are given by
\begin{equation}
\label{eq:populations}
\begin{aligned}
\begin{array}{ll}
p_{\mathrm{correct}} =
(1-\varepsilon_2)\!\Big(1-\tfrac{(s-1)\varepsilon_1}{s}
\Big) + \tfrac{\varepsilon_2}{d} &\quad\text{($1$ state)}, \\[4pt]
p_{\mathrm{in\text{-}block}} = (1-\varepsilon_2)\tfrac{\varepsilon_1}{s} + \tfrac{\varepsilon_2}{d} 
&\quad\text{($s-1$ states)}, \\[4pt]
p_{\mathrm{cross\text{-}block}} = \tfrac{\varepsilon_2}{d} 
&\quad\text{($d-s$ states)}.
\end{array}
\end{aligned}
\end{equation}

Consider a signal subspace \(\mathscr S_b\) with projector \(P_b\) and signals \(\{\ket{\psi_{b,x}}\}_{x=0}^{k-1}\).
For each signal \(\ket{\psi_{b,x}}\), the conditional kept probability is
\begin{align}
\label{eq:alpha-per-signal}
&\Pr[\text{kept}\mid \text{basis match}, b,x]
\nonumber\\
&
=(1-\varepsilon_2)\Big[(1-\varepsilon_1)\,\operatorname{Tr}\!\big(|\psi_{b,x}\rangle\langle\psi_{b,x}|\,P_b\big) \nonumber\\
&\quad\ + \varepsilon_1\,\operatorname{Tr}\!\big(\Phi_{\mathrm{block}}^{(b)}(|\psi_{b,x}\rangle\langle\psi_{b,x}|)\,P_b\big)\Big]
+ \varepsilon_2\,\frac{k}{d}
\end{align}
for every given $x\in \{0,1,\ldots,k-1\}$
(see \eqref{Equation:Phi_b} for the definition of $\Phi_{\mathrm{block}}^{(b)}$).

Averaging over $x\in \{0,\ldots,k-1\}$, we obtain the following expression for   the  kept probability:
\begin{equation}
\label{eq:alpha-avg}
\alpha_{b,\mathfrak B} 
=(1-\varepsilon_2)\!\left(1-\frac{(s-1)\varepsilon_1}{s}+\frac{\varepsilon_1}{s}(E_b-1)\right)
+\varepsilon_2\frac{k}{d}
\end{equation}
where we have defined $E_b$ as the block overlap:
\begin{equation}
\label{eq:Eb-def}
\begin{gathered}
E_b := \frac{s}{k}\sum_{x=0}^{k-1}\operatorname{Tr}\!\big(\Phi_{\mathrm{block}}^{(b)}(|\psi_{b,x}\rangle\langle\psi_{b,x}|)\,P_b\big) \\
\text{with } 1 \le E_b \le s.
\end{gathered}
\end{equation}
The quantity \(E_b\) measures the degree to which the smeared weight remains inside the kept subspace \(\mathscr S_b\) on average. If \(E_b\) takes a large value, this means that the encoding is block-compatible with \(P_b\), hence intra-block mixing leaves most probability in \(\mathscr S_b\). Whereas if \(E_b\) is small, the error tends to propagate outside \(\mathscr S_b\).

The corresponding dit error within the kept subspace is thus
\begin{equation}
\label{eq:Q-bbc-final}
Q_{b,\mathfrak B} 
=\frac{\Big((1-\varepsilon_2)\frac{\varepsilon_1}{s}+\frac{\varepsilon_2}{d}\Big)(E_b-1)
+\frac{\varepsilon_2}{d}\,(k-E_b)}{\alpha_{b,\mathfrak B}}.
\end{equation}

\paragraph{Optimal encoding.}
For a basis-aligned truncation, let the kept projector \(P_b\) select \(\ell_m\) computational states from block \(m\), with \(\sum_{m=0}^{s-1} \ell_m = k\).
By \eqref{eq:Eb-def}, the block-overlap reduces to
\begin{equation}
E_b \;=\; \frac{1}{k}\sum_{m=0}^{s-1} \ell_m^2.
\end{equation}
To minimize \(E_b\) at a fixed \(k\), we balance the occupancy across the blocks: set $
k \;=\; s\,q + t, \quad q=\big\lfloor k/s \big\rfloor,\quad t\in\{0,\ldots,s-1\},$
where \(t\) is the remainder (not to be confused with the in–block index \(r\)). 
Assign \(\ell_m\in\{q,q+1\}\) such that exactly \(t\) blocks take \(\ell_m=q+1\) and the remaining \(s-t\) blocks take \(\ell_m=q\). This yields the global minimum
\begin{equation}
E_{\min} \;=\; \frac{s q^2 + 2 q t + t}{k}.
\end{equation}
Substituting \(E_b=E_{\min}\) into \eqref{eq:alpha-avg} and \eqref{eq:Q-bbc-final} yields the optimal kept probability \(\alpha^{\text{th}}_{\mathfrak B}\) and dit error \(Q^{\text{th}}_{\mathfrak B}\) for basis-aligned encodings under the block-biased channel.

\begin{figure*}[!ht]
    \centering
    \includegraphics[width=1\columnwidth]{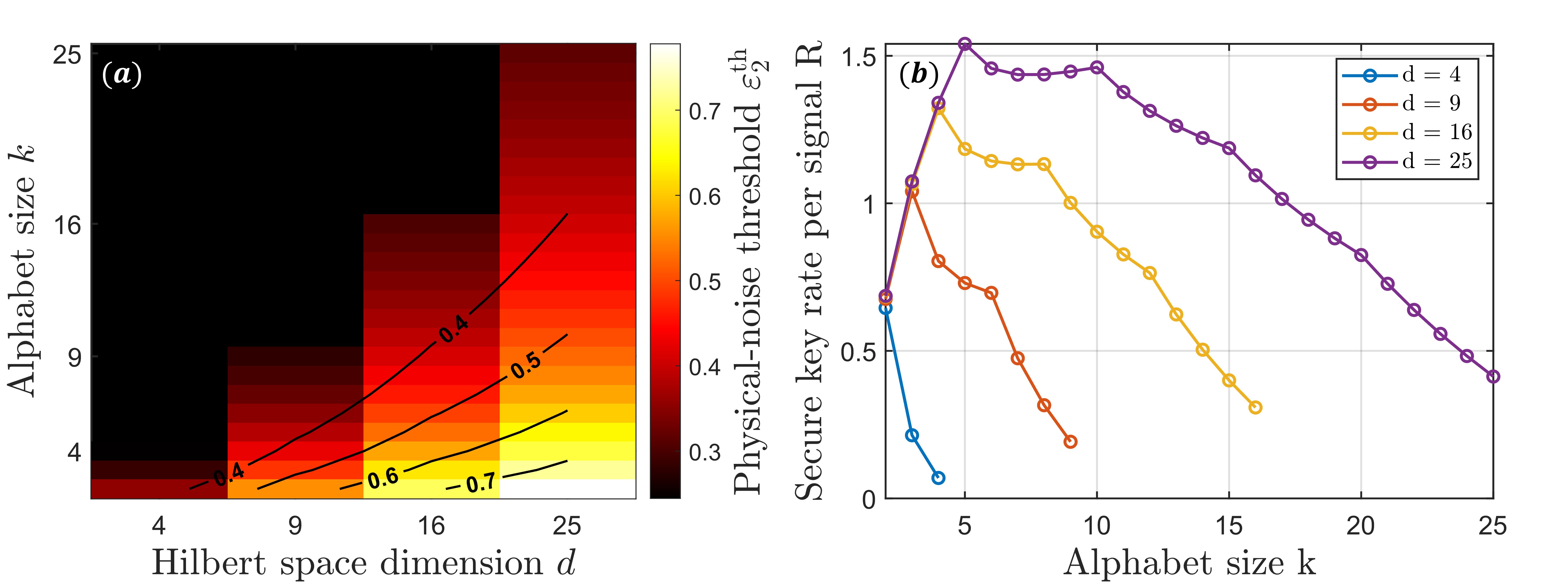}
        \caption{\textbf{Physical-noise threshold and secure key rate for the block-bias channel.} 
    \textbf{(a) Heatmap of the physical-noise threshold $\varepsilon_{2}^{\mathrm{th}}$ as a function of the Hilbert space dimension $d$ and signal-set size 
    $k$, with intra-block depolarization fixed at $\varepsilon_{1}=0.07$.} The contour lines highlight threshold levels, showing how the tolerance to inter-block noise depends on both $d$ and the chosen signal-set. 
    \textbf{(b) Secure key rate $R$ as a function of signal-set size $k$, for a space  dimension of $d=4,9,16,25$.} The optimal signal-set size is $k_{\text{opt}}=\sqrt{d}$, where $\varepsilon_1=0.3$ and $\varepsilon_2=0.07$.}
    \label{fig:pth_vs_k_block}
\end{figure*}

As illustrated in Fig.~\ref{fig:coding_blockfig}(b), for \(s=3\) the optimal encoding distributes the chosen states as evenly as possible among the three blocks, thereby minimizing the number of selected states within the same block and reducing within block confusions. In this example, for the dimensions \(s=3\), \(d=9\), and the noise parameters $\varepsilon_1=0.3$, $\varepsilon_2=0.07$, we will see (Fig. \ref{fig:pth_vs_k_block}(b)) that the optimal encoding is for \(k=3\). Specifically, one may choose $\mathscr{S}_Z=\{\ket{1},\ket{4},\ket{7}\}$
and $\mathscr{S}_X=\{\ket{\mu_1},\ket{\mu_4},\ket{\mu_7}\}$.

\paragraph{Key rate and error thresholds.}
\label{subsec:DW+thresholds}
For the basis alignment channel, the errors are symmetric, i.e., $Q_{X,\mathfrak B}=Q_{Z,\mathfrak B}=Q_{\mathfrak B}$, and the asymptotic Devetak-Winter bound per sifted symbol is: 
\begin{equation}
\begin{aligned}
R_{\mathrm{per\text{-}sifted\text{-}symbol}} &\ge \log_2 k - 2\,h_k(Q_{\mathfrak B}), 
\end{aligned}
\end{equation}
where \(h_k(\cdot)\) is as in \eqref{eq:hk}.

The $k$-ary threshold, $Q^{\text{th}}_{\mathfrak B}$, solves
\begin{equation}
h_k(Q^{\text{th}}_{\mathfrak B})=\tfrac12\log_2 k.
\end{equation}
The threshold $Q^{\text{th}}_{\mathfrak B}$ can also be expressed as in \eqref{eq:Q-bbc-final}, taking $E_b=E_{\min}$.

Setting $Q_{\mathfrak B}=Q_{\mathfrak B}^{\text{th}}$ yields the inter-block noise threshold
\begin{equation}
\varepsilon_{2,\mathfrak B}^{\mathrm{th}}(\varepsilon_1)
= \frac{Q_{\mathfrak B}^{\text{th}}  \,K_0 - C_0}{\,C_1 - Q_{\mathfrak B}^{\text{th}}  (-K_0 + k/d)},
\end{equation}
with
\begin{equation}
\begin{alignedat}{2}
  K_0 &\,=\, 1-\tfrac{(s-1)\varepsilon_1}{s}+\tfrac{\varepsilon_1}{s}(E_{\min}-1),\\
  C_0 &\,=\, \tfrac{\varepsilon_1 (E_{\min}-1)}{s},\\
  C_1 &\,=\, -\tfrac{\varepsilon_1 (E_{\min}-1)}{s}+
  \tfrac{k-1}{d}.
\end{alignedat}
\end{equation}

Figure~\ref{fig:pth_vs_k_block}(a) depicts the physical-noise threshold. For a uniform basis selection, the sifting rate is $\tfrac12\,\alpha_{\mathfrak B}$.  
The Devetak–Winter bound then gives
\begin{equation}
\label{eq:key-bb}
R_{\text{per-signal}}
\;\ge\;
\frac12\,\alpha_{\mathfrak B}\,
\Big[\log_2 k - 2\,h_k(Q_{\mathfrak B})\Big],
\end{equation}
See Figure~\ref{fig:pth_vs_k_block}(b).
We note that the zero rate is obtained exactly at the threshold value, $\varepsilon_{2,\mathfrak B}^{\mathrm{th}}$.

\paragraph{Advantage of reduced state embedding}
Figure~\ref{fig:pth_vs_k_block}(b) highlights the advantage of our reduced state embedding scheme for the block-biased channel. The figure depicts the key rate $R_{\text{per-signal}}$ as a function of the signal-set size $k$, for a space  dimension of  $d=4,9,16,25$. For noise parameters $\varepsilon_1=0.3$ and $\varepsilon_2=0.07$ the optimal signal-set size turns out to be $k=s=\sqrt{d}$. As before, this confirms  that encoding into a reduced subspace is preferable.

\begin{figure*}[!ht]
    \centering
    \includegraphics[width=0.7\columnwidth]{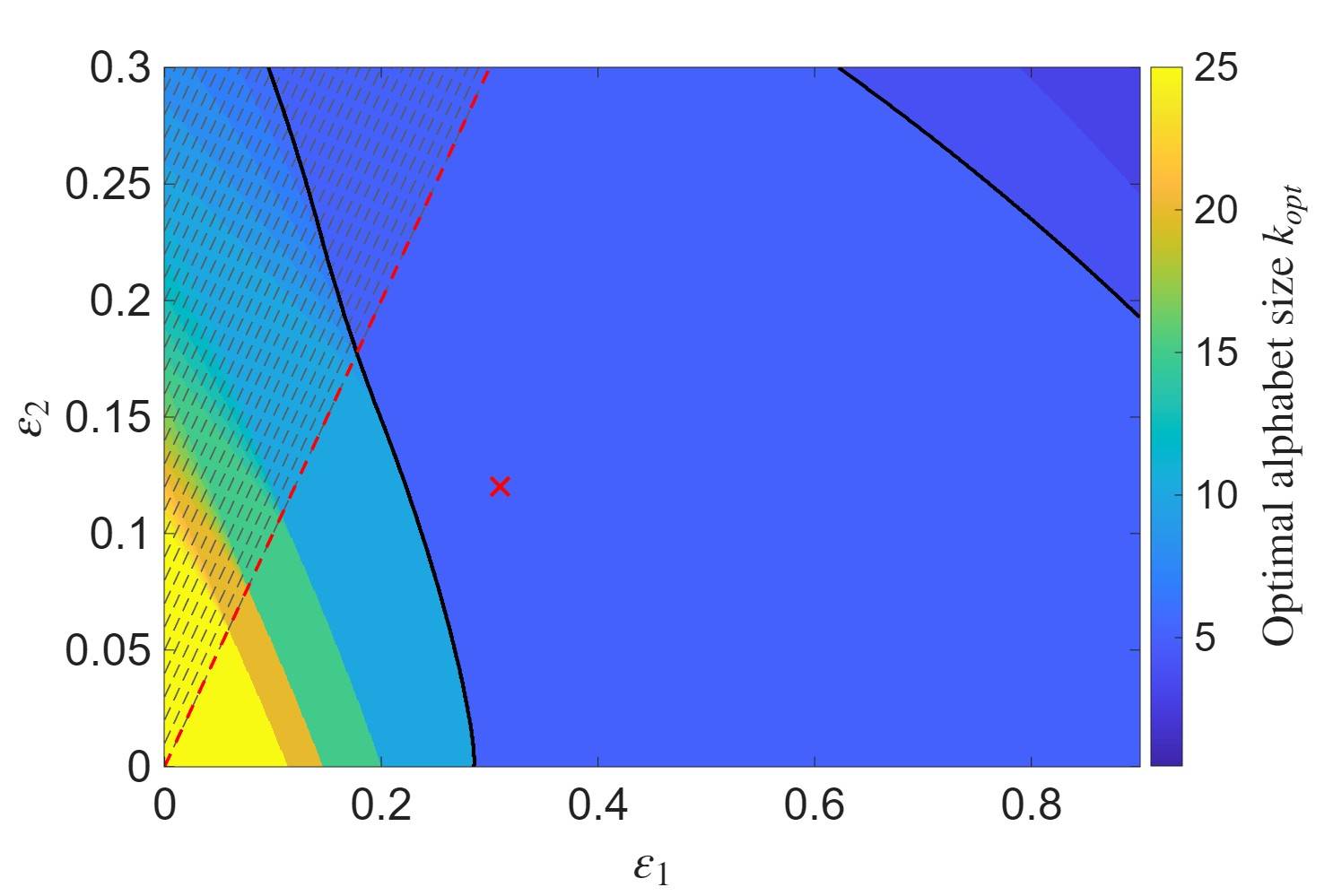}
        \caption{\textbf{Heatmap of the optimal subspace dimension $k_{\mathrm{opt}}$ within a Hilbert space of dimension $d=25$, shown as a function of the two noise parameters $(\varepsilon_1,\varepsilon_2)$.}  
    The block size is $s=5$. The color scale indicates the value of the optimal signal-subspace dimension, showing that the preferred encoding changes across the noise-parameter space. The red dashed line marks the boundary between the block-biased regime, ($\varepsilon_1>\varepsilon_2$), and the regime that is no longer block-biased. The hatched region indicates the excluded parameter space, $\varepsilon_1\le\varepsilon_2$. The red cross marks the fitted experimental noise parameters in Section~\ref{sec:validation}.}
    \label{fig:epsilon_k}
\end{figure*}

In general, the optimal scheme depends on the underlying noise model. As shown above, different models, such as depolarizing, modulo, and block-biased channels, lead to different signal constellations. Moreover, even within a fixed noise model, here the block-biased model, the noise level (i.e., the specific values of the channel parameters) affects the optimal signal set size 
$k_{\text{opt}}$, which decreases as the noise increases.  Figure~\ref{fig:epsilon_k} depicts 
 $k_{\text{opt}}$ as a function of the parameters 
$(\varepsilon_1,\varepsilon_2)$, in the regime 
$\varepsilon_1>\varepsilon_2$. The heat map shows that the optimal encoding changes with the noise parameters.

\begin{remark}
    In some implementations, block bias arises from the state arrangement and not the measurement, hence it appears in only one basis. For example, in multi-mode fibers, modes with similar propagation constants couple more strongly.
    If the block structure is fixed in the computational basis \(Z\) (e.g., a basis-anchored map \(\Phi_{\mathrm{block}}^{(Z)}\)), and we encode/measure in a different MUB \(X\), the error pattern in \(X\) is not block-biased: the \(Z\)-localized mixing spreads across many \(X\)-eigenstates. In this cross-basis situation the relevant overlap becomes $E_{X}^{(Z)}\;:=\; \frac{s}{k}\sum_{t=0}^{k-1}\Tr\!\big(\Phi_{\mathrm{block}}^{(Z)}(|\psi_{X,t}\rangle\langle\psi_{X,t}|)\,P_X\big)$.
    The kept probability and dit error retain the same expression forms as in \eqref{eq:alpha-avg}–\eqref{eq:Q-bbc-final}, with \(E_b\) replaced by \(E_{X}^{(Z)}\).
    Operationally, this yields a different dit errors in \(X\) than in \(Z\), because the confusion matrix in \(X\) lacks block structure and exhibits broader spreading.
\end{remark}

\subsection{Experimental Validation in a 25-dimensional QKD System}
\label{sec:validation}

We validate reduced‑state embedding in an entanglement‑based QKD platform by fixing the Hilbert‑space dimension at $d{=}25$ and, for each $k\in \{2,\ldots,25\} $, implementing an embedded $k$‑dimensional signal set in both mutually unbiased bases (MUBs). Bob’s measurement uses the $(k{+}1)$‑outcome filter of Eq.~\eqref{eq:POVM}; from basis‑matched, conclusive coincidences we estimate the kept‑event probability $\alpha$ and the conditional $k$‑ary dit error $Q$, and we compute the secure key rate per signal using Eq.~\eqref{eq:key-bb}. For each $k$ we choose the optimal embedded $k$‑dimensional subspace Section~\ref{Subsection:block}.

The physical system distributes spatially entangled photon pairs in a $5{\times}5$ pixel basis ($d{=}25$) and realizes two MUBs with a multi‑plane light converter (MPLC), namely a cascade of phase planes separated by free‑space propagation that implements programmable unitary mode sorting on spatial modes. In this architecture the two MUBs are effected by applying five‑point discrete Fourier transforms (DFTs) along rows or columns of the grid, which yields the block‑biased error structure modeled in Section~\ref{Subsection:block}. Complete experimental details are provided in Methods~\ref{sec:methods-exp-details-43}. The same MPLC platform previously demonstrated a $d{=}25$ QKD protocol with two MUBs; here we extend its capability by validating arbitrary embeddings with $k<d$ in the same $d$‑dimensional Hilbert space.

For each $k\in \{2,\ldots,25\} $ and for each basis $b\in{Z,X}$, both parties measure their halves of each entangled pair in randomly chosen bases, and only basis‑matched, conclusive coincidences contribute to the $k$‑ary data. The quantities $\alpha$ and $Q$ are obtained from the data using Eqs.~\eqref{eq:alpha-avg}–\eqref{eq:Q-bbc-final}, and the secure key rate per signal follows from Eq.~\eqref{eq:key-bb}.

\begin{figure*}[!ht]
\centering
\includegraphics[width=0.8\columnwidth]{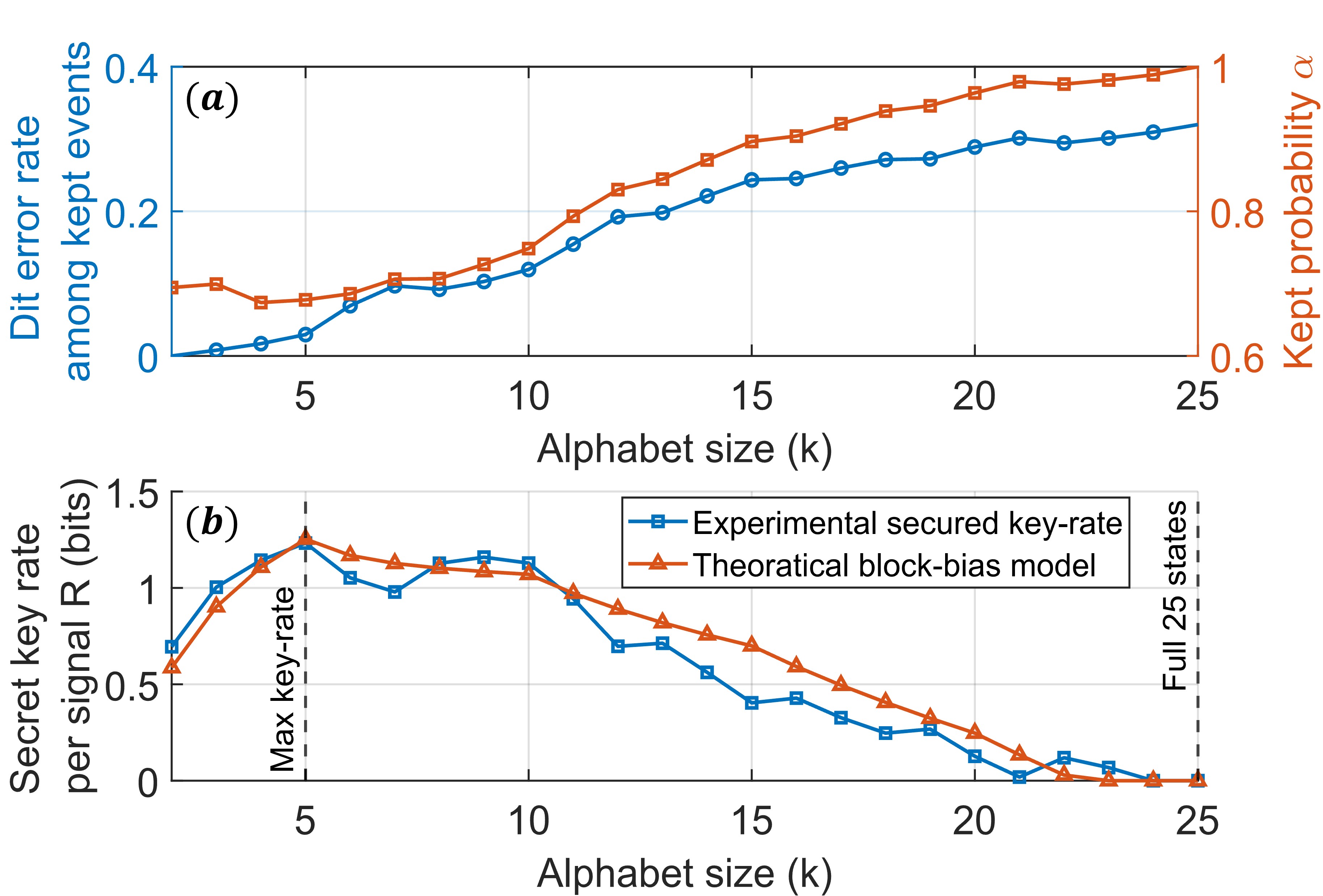}
\caption{\textbf{Validation in $d{=}25$ with reduced‑state embedding.} \textbf{(a)} Dit error $Q$ among kept events and kept‑event probability $\alpha$ versus embedded dimension $k$. \textbf{(b)} Secure‑key rate per signal, evaluated via Eq.~\eqref{eq:DW} with parameters extracted using Eqs.~\eqref{eq:alpha-avg}–\eqref{eq:Q-bbc-final}, exhibits a clear maximum at $k{=}5$. The trends are consistent with the block‑bias analysis of Section~\ref{Subsection:block}, for $\varepsilon_1=0.31$ and $\varepsilon_2=0.12$, where increasing $k$ raises payload but also channels more physical noise into conclusive outcomes, producing a noise‑dependent interior optimum with $k<d$.}
\label{fig:kfir_theo}
\end{figure*}

Figure~\ref{fig:kfir_theo} shows that the kept event probability $\alpha$ and the dit error rate among kept event $Q$ increase with $k$, while the secure‑key rate has a pronounced maximum at $k=5$. This optimum, obtained for the chosen optimal embedded subspace at each $k$, agrees with Section~\ref{Subsection:block} and validates that, on this entanglement‑based $d{=}25$ platform, operating in an embedded $k$‑dimensional subspace outperforms using all $d$ states under realistic block‑biased noise.

\section{DISCUSSION}

We introduce the method of reduced
state embeddings to quantum key distribution (QKD): a $k$-dimensional signal set embedded in a $d$-dimensional Hilbert space, where $k<d$. The idea balances two effects. Using more modes increases the information one can extract from each successful detection (conclusive event). However, restricting to a smaller, ``cleaner" subsets screens out physical noise before reconciliation through the kept probability $\alpha$ and the conditional dit error $Q$.
This trade-off produces an interior optimum $k_{\text{opt}}$: Beyond this point, adding modes, such that $k>k_{\text{opt}}$, creates more in-subset confusion than benefit, and the secure key rate drops.
In our demonstration, the experimental maximum is attained at \(k_{\text{opt}}=5\) for \(d=25\). The optimality of a strictly reduced state embedding results from the competition between these factors. See Section~\ref{Subsection:block} for the theoretical analysis and Section~\ref{sec:validation} for the experimental validation.

Channel structure dictates both the subset size \(k\) and the symbol geometry. For isotropic depolarization (see 
Section~\ref{Subsection:Dep}), a small physical error \(\varepsilon\) favors covering the full dimension with (\(k=d\)), while above a noise-dependent crossover, an intermediate \(k<d\) yields a higher per-signal rate due to a reduced dit error \(Q  \) under post-selection. For nearest-neighbor cross-talk modulo channel, the adjacency count \(W(S)\) is the key design variable: maximally spaced sets with no adjacencies suppress kept errors and yield the threshold value of the physical error, \(\varepsilon^{\text{th}}=1/2\), at the cost of a lower throughput (see Section~\ref{Subsection:Mod}). 
In contrast, (strictly) reduced state embedding is suboptimal for channels without symbol confusion, e.g., flagged erasure or dephasing in the key basis. That is, in such cases, the optimal signal-set is the entire bases, as $k_{\text{opt}}=d$.
In general, the optimal scheme depends on the underlying noise model. Different noise models, such as depolarizing, modulo, and block-biased channels, lead to different signal constellations. Moreover, even within a fixed noise model, the noise level affects the optimal signal set size 
$k_{\text{opt}}$, which decreases as the noise increases (see Figure \ref{fig:epsilon_k} for the block-biased channel). 
 
Conceptually, reduced-state embeddings echo Shannon’s classical error-correction approach \cite{shannon_mathematical_1948}, but it is not identical. Specifically, for a modulo channel, a reduced-state embedding can guarantee strictly zero error, in a similar manner as in Shannon's zero-error codes based on confusability graphs~\cite{shannon_zero-error_1956}. Here, however, we focus on quantum state embedding in the QKD setting, and embed a \(k\)-ary code inside a larger \(d\)-dimensional Hilbert space, and then use a conclusive/inconclusive measurement such that part of the physical noise is actively filtered into inconclusive outcomes. 

The scheme introduced here lies at the intersection of what are traditionally referred to as encoding, modulation, and error correction~\cite{viterbi2009principles,proakis2001digital}, and no single term fully captures its role. On the one hand, the construction embeds logical states into a higher-dimensional Hilbert space with deliberate structure and redundancy, leading to enhanced distinguishability and robustness to noise. In this sense, our methods share important features with error-correcting encodings~\cite{nielsen2010quantum}. On the other hand, it does not rely on explicit recovery operations, such as syndrome extraction, and therefore does not constitute error correction in the conventional coding-theoretic sense~\cite{gottesman1997stabilizer}. At the same time, describing the scheme as modulation alone is also incomplete, as the mapping is discrete and structured rather than a standard continuous signal representation. This ambiguity reflects a broader conceptual overlap, already present in quantum computing settings such as GKP encodings~\cite{gottesman2001encoding}, where representation and noise protection are intertwined~\cite{terhal2015quantum}. More generally, in quantum communication the functions of state encoding, physical-layer modulation and detection, and error protection are often more tightly coupled to the underlying hardware than in classical layered architectures, and can therefore be co-designed at the level of the physical implementation~\cite{djordjevic2019physical,koudia2025physical}. 

In the framework of quantum error correction (QEC) for quantum computing, our reduced-state embedding realizes an explicit erasure-type error-detection step within the quantum channel. Lo and Chau first formulated QKD security in terms of entanglement purification and quantum error-correcting codes~\cite{lo1999unconditional}, and Shor and Preskill subsequently showed that the same protection can be achieved through classical post-processing~\cite{shor2000simple}. Consequently, most QKD implementations apply error correction only after measurement, without physical QEC on the quantum states themselves. We realize a practical encoding scheme at the physical layer: a \(k\)-ary logical alphabet is embedded in a \(d\)-dimensional space, and Bob’s \((k{+}1)\)-outcome filter acts as a syndrome test that converts physical errors into erasures before post-processing. We show that this QEC erasure conversion strengthens the tolerance to noise and improves key rates under realistic high-dimensional noise.
    
An analogous trade-off appears in quantum computing, where erasure-based conditioning enhances gate fidelity by selectively retaining high-purity outcomes~\cite{baranes2025leveraging}. Heralded entangling gates in photonic platforms or Rydberg-atom arrays exploit the same principle~\cite{ma2023high,scholl2023erasure}: a noisy physical interaction is projected onto a nearly unitary subspace when measurement or loss events are treated as erasures rather than logical errors. In all these systems, fidelity improves because throughput is traded for conditional purity. Our reduced-state embedding operates on the same logic where the effective channel seen by the Devetak–Winter process is thereby purified, producing an interior optimum in the kept signal-set size $k$ analogous to the conditional-fidelity optimum in two-qubit operations based on erasure. This correspondence highlights a unifying idea, selective erasure as a route to higher logical fidelity, spanning both quantum communication and computation.
    
Finally, because receiver-side filtering is agnostic to the physical realization of the \(d\)-dimensional space, the method is inherently scalable. It portably extends across spatial families (e.g., orbital angular momentum), non-spatial degrees of freedom (temporal or frequency encodings), and photon statistics (heralded single photons~\cite{schiavon2016heralded}, or decoy-state coherent pulses~\cite{lo_decoy_2005}), provided two MUBs and compatible sorting or projective filters are available. This establishes reduced-state embedding as a practical, noise-resilient resource for scalable quantum cryptography.

\section{METHODS}
\subsection{Experimental details}
\label{sec:methods-exp-details-43}

The MPLC-based spatial-mode QKD platform was previously reported in Ref.~\cite{lib_high-dimensional_2025}; for completeness, we briefly summarize here the essential parameters required to implement the system and to carry out our validation in 
Section~\ref{sec:validation}.

\textit{Source and state preparation.} Spatially entangled photon pairs are generated via type‑I SPDC in an 8\,mm BBO crystal pumped by a 405\,nm continuous‑wave laser (Cobolt 06‑MLD). The pump power is 125\,mW and reduced to $\sim$30\,mW to limit accidentals within a 400\,ps coincidence window. A $f{=}150$\,mm lens images the far‑field onto a binary amplitude mask comprising 50 circular apertures (radius 100\,$\mu$m), defining a 25‑dimensional pixel basis.

\textit{Multi‑plane light converter (MPLC).} The photon pair coherent manipulation is obtained by a 10‑plane MPLC by bouncing the photon ten times between a reflective phase SLM (Hamamatsu X13138‑02) and a mirror. For each basis, ten $140{\times}360$ pixel phase masks are computed by wavefront matching (30 iterations). The two MUBs used for $d{=}25$ are realized by applying $\sqrt d{=}5$‑point DFTs along rows (MUB 1) or columns (MUB 2) of the $5\times5$ grid, so only five modes interfere at a time, reducing optical depth and enforcing the block‑biased error geometry.

\textit{Detection and timing.} Correlations are recorded with two 100\,$\mu$m‑core fibers coupled to single‑photon avalanche diodes (Excelitas SPCM‑AQRH‑62‑FC) and time‑tagged (Swabian Instruments Time Tagger 20). The detection plane is 43.5\,mm after the last MPLC plane. For each sent state, coincidence counts are integrated for 100\,s and normalized to obtain conditional probabilities; 20\,nm‑wide bandpass filters are placed before the detectors, and a dichroic mirror removes the pump. A foldable beam splitter before the MPLC allows direct measurements in the computational basis when required. 

\textit{Losses.} Total MPLC loss depends on the programmed transformation. Average insertion loss per photon is $\sim$10.7\,dB for the $d{=}5$ configuration and $\sim$13.4\,dB for $d{=}25$, estimated from coincidence rates before and after the MPLC. Using static (non‑programmable) phase plates would further reduce loss but at the expense of reconfigurability.

\section*{RESOURCE AVAILABILITY}
\subsection*{Lead contact}
Requests for further information and resources should be directed to and will be fulfilled by the lead contact, Uzi Pereg (uzipereg@technion.ac.il)

\subsection*{Materials availability}
The data that supports the findings of this study is available from the corresponding author upon reasonable request. The experimental data used to validate our protocol in section~\ref{sec:validation} is available in Ref.~\cite{lib_data_2024}.

\subsection*{Data and code availability}
Any additional information required to reanalyze the data reported in this paper is available from the lead contact upon request.

\section*{ACKNOWLEDGMENTS}
The authors thank Ohad Lib and Yaron Bromberg for conducting the experiments used to validate our protocol, and thank Guy Bartal, Meir Orenstein, Aviv Karnieli, and Avi Zadok for fruitful discussions. A.K. acknowledges the support from the Azrieli fellowship and the support from the Helen Diller Quantum Center at the Technion. K.S acknowledges the support of the Israeli Council for Higher Education, the Zuckerman STEM Leadership Program, and the European Union’s Horizon program under Marie Skłodowska-Curie grant No. 101202109. S.T. acknowledges support from the Rothschild fellowship of the Yad Hanadiv foundation, the VATAT Quantum fellowship of the Israel Council for Higher Education, the Helen Diller Quantum Center postdoctoral fellowship and the Viterbi fellowship of the Technion - Israel Institute of Technology.
This research was supported by the Israel Science Foundation (ISF), Grants 939/23 and 2691/23, German-Israeli Project Cooperation (DIP) within the Deutsche Forschungsgemeinschaft (DFG), Grant 2032991, the Junior Faculty Program for Quantum Science and Technology of the Planning and Budgeting Committee of the Council for Higher Education of Israel through Grant 86636903.

\section*{AUTHOR CONTRIBUTIONS}
A.K., K.S., and S.T. conceived the project. A.K. and K.S. performed the noise-model analysis and derived the secure key-rate bounds under truncation. U.P. supervised the research. All authors discussed the results and participated in writing the manuscript.

\section*{DECLARATION OF INTERESTS}
The authors declare no competing interests.

\bibliography{references}

\section*{Tables}
\begin{table}[ht]
\centering
\begin{tabular}{ll}
\hline
\textbf{Parameter} & \textbf{Notation} \\
\hline
Hilbert space dimension & $d$ \\
Subspace dimension & $k$ \\
Optimal dimension & $k_{\text{opt}}$ 
\\
Full Z-basis & $\mathcal B_Z =\{\ket{0},\ldots,\ket{d-1}\}$
\\
Full X-basis & $\mathcal B_X  =\{\ket{\mu_0},\ldots,\ket{\mu_{d-1}}\}$\\
Z-basis subset (simple) & $\mathsf S_Z=\{\ket{0},\ldots,\ket{k-1}\}$ \\
X-basis subset (simple) & $\mathsf S_X=\{\ket{\mu_0},\ldots,\ket{\mu_{k-1}}\}$ \\
Z-basis subset (general) & $\mathscr S_Z=\{\ket{\phi_0},\ldots,\ket{\phi_{k-1}}\}$ \\
X-basis subset (general) & $\mathscr S_X=\{\ket{\sigma_0},\ldots,\ket{\sigma_{k-1}}\}$\ 
\\
Symbol& 
$x\in \{0,1,\ldots,k-1\}$
\\
Basis & $b\in\{X,Z\}$
\\
Transmitted state & $\ket{\psi_{b,x}}$\\
Conclusive outcome & $\Pi_{b,x}$\\
Inconclusive outcome & $\Pi_{b,\perp}$\\
Signal subspace projector & $P_b=\sum_{x=0}^{k-1}\Pi_{b,x}$\\
\hline
\end{tabular}
\caption{\textbf{Notation table 1.} List of notations used in Section~\ref{definitions}.}
\label{tab:notation_1}
\end{table}

\begin{table}[H]
\centering
\begin{tabular}{lccc}
\hline
\textbf{Notation} & \textbf{Depolarizing} & \textbf{Modulo} & \textbf{Block} \\
\hline
Channel & $\mathfrak{D}$ & $\mathfrak{M}$ & $\mathfrak{B}$
\\
Channel parameters & $\varepsilon$ & $\varepsilon$ & $\varepsilon_1$,  $\varepsilon_2$
\\
Kept probability ($b$-basis)& $\alpha_{b,\mathfrak D}$ & $\alpha_{b,\mathfrak M}$ & $\alpha_{b,\mathfrak B}$ \\
 Quantum error rate ($b$-basis) & $Q_{b,\mathfrak D}$ & $Q_{b,\mathfrak M}$ & $Q_{b,\mathfrak B}$ \\
 Kept probability (symmetric) & $\alpha_{\mathfrak D}$ & $\alpha_{\mathfrak M}$ & $\alpha_{\mathfrak B}$ \\
Quantum error rate (symmetric) & $Q_{\mathfrak D}$ & $Q_{\mathfrak M}$ & $Q_{\mathfrak B}$
\\
Parameter threshold & $\varepsilon_{\mathfrak D}^{\text{th}}$ & $\varepsilon_{\mathfrak M}^{\text{th}}$ & $\varepsilon_2^{\text{th}}(\varepsilon_1)$
\\
Kept probability threshold & $\alpha_{\mathfrak D}^{\text{th}}$ & $\alpha_{\mathfrak M}^{\text{th}}$ & $\alpha_{\mathfrak B}^{\text{th}}$
\\
Quantum error threshold & $Q_{\mathfrak D}^{\text{th}}$ & $Q_{\mathfrak M}^{\text{th}}$ & $Q_{\mathfrak B}^{\text{th}}$ \\
\hline
\end{tabular}
\caption{\textbf{Notation table 2.} List of channel-specific notations used in Section~\ref{sec:performance}.}
\label{tab:notation_2}
\end{table}

\newpage
\section*{Figures}

\begin{figure*}[!ht]
    \centering
    \includegraphics[width=0.6\columnwidth]{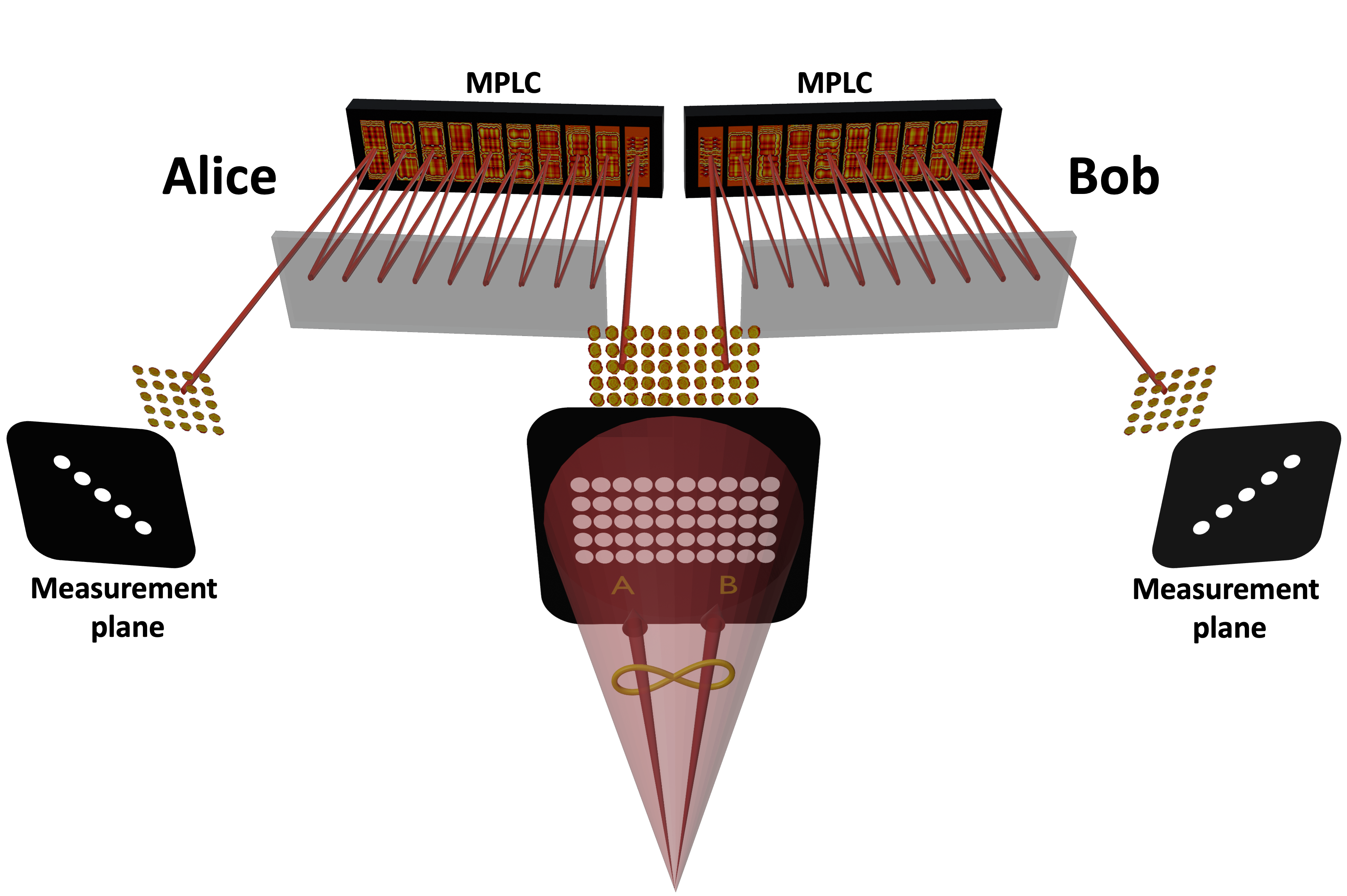}
    \caption{\textbf{Illustration of system under test (}$d{=}25$\textbf{):} entangled photon pairs are filtered to a $5\times5$ pixel basis and routed to two 10‑plane MPLCs. The two MUBs are realized, for example, by applying five-mode DFTs along rows or columns. Basis choice is performed by switching displayed phase masks. This construction yields block‑biased errors consistent with Section~\ref{Subsection:block}. Two additional masks at the measurement plane, implemented digitally, represent the dimension reduction from $d=25$ to $k=5$.}
    \label{fig:setup_hd}
\end{figure*}

\end{document}